\documentclass{desyproc}
\begin{document}
%------------------------------------		
\title{Analyzing New Physics in  
\boldmath{$\bar{B}^0 \to D^{(\ast)}\tau^-\bar\nu_{\tau}$}}
	
%for single authors the superscripts are optional
\author{{\slshape C.~T.~Tran$^{1,2}$, M.~A.~Ivanov$^1$, J.~G.~K\"{o}rner$^3$}\\[1ex]
$^1$Bogoliubov Laboratory of Theoretical Physics JINR, 141980 Dubna, Russia\\
$^2$Moscow Institute of Physics and Technology, 141700 Dolgoprudny, Russia\\
$^3$PRISMA Cluster of Excellence, Institut f\"{u}r Physik, 
Johannes Gutenberg-Universit\"{a}t, 
D-55099 Mainz, Germany}

% please enter the contribution ID for the DOI
\contribID{xy}

% TO THE CONFERENCE EDITORS: 
% please update the following information      
% before sending the template to the authors
\confID{999}  % if the conference is on Indico uncomment this line
\desyproc{DESY-PROC-2099-01}
\acronym{VIP2010} % if you want the Acronym in the page footer uncomment this line
\doi  % if there is an online version we will register DOIs

\maketitle

\begin{abstract}
We discuss possible new physics (NP) effects beyond the standard model (SM) in the exclusive decays
$\bar{B}^0 \to D^{(\ast)} \tau^- \bar{\nu}_{\tau}$. Starting with a model-independent effective Hamiltonian including non-SM
  four-Fermi operators, we show how to obtain experimental constraints on different NP
  scenarios and investigate their effects on a large set of physical observables. The $\bar{B}^0 \to D^{(\ast)}$ 
transition form factors are calculated in the full kinematic $q^2$ range 
by employing the covariant confined quark model developed by our group. 
\end{abstract}

\section{Introduction}
The exclusive semileptonic decays $\bar{B}^0 \to D^{(\ast)}\tau^-\bar\nu_{\tau}$ have
been measured by the {\it BABAR}~\cite{Lees:2012xj}, Belle~\cite{Belle}, and LHCb~\cite{Aaij:2015yra} collaborations in an effort to
unravel the well-known $R_{D^{(*)}}$ puzzle  which has persisted for several
years (see~\cite{Ivanov:2015tru,Ivanov:2016qtw,Ivanov:2017mrj} and references therein). The current
world averages of the ratios are $R_D = 0.406 \pm 0.050$ and
$R_{D^\ast} = 0.311 \pm 0.016$, which exceed the SM predictions of
$R_D = 0.300 \pm 0.008$~\cite{Na:2015kha} and $R_{D^\ast}
= 0.252 \pm 0.003$~\cite{Fajfer:2012vx} by $2.1\sigma$ and $3.6\sigma$, respectively.

The excess of $R(D^{(\ast)})$ over SM predictions has attracted a great deal
of attention in the particle physics community and has  led to many theoretical
studies looking for NP explanations. Some studies focus on specific NP models
including two-Higgs-doublet models~\cite{2HDMs}, leptoquark
models~\cite{leptoquark}, and other extensions of the SM. Other studies adopt a model-independent approach, in which a
general effective Hamiltonian for the $b \to c \ell \nu$ transition in the
presence of NP is imposed to investigate the impact of various NP operators
on different physical observables~\cite{4FermiO}.
Most of the theoretical studies rely on the heavy quark effective theory
(HQET)~\cite{Neubert:1993mb} to evaluate the hadronic form
factors, which are expressed through a few universal functions in the heavy
quark limit (HQL). In the present analysis, we employ an alternative
approach to calculate the NP-induced hadronic transitions
based on the covariant confined quark model (CCQM), which has
been developed in some earlier papers by us
(see~\cite{CCQM} and references therein). 

Here we follow the authors of~\cite{4FermiO} to include NP operators in the effective Hamiltonian and investigate their effects on physical observables of the decays $\bar{B}^0 \to D^{(\ast)} \ell^-\bar\nu_{\ell}$. We define a full set of form factors corresponding to SM+NP operators and calculate them by employing the CCQM. In the CCQM the
transition form factors can be determined in the full range of  momentum
transfer, making the calculations straightforward without any extrapolation. This provides an opportunity to investigate NP operators in a self-consistent manner, and independently from the HQET. We first constrain the NP operators using experimental data on the ratios of branching fractions, then analyze their effects on various observables. We also derive the fourfold angular distribution for the cascade decay $\bar {B}^0\to D^{\ast\,+}(\to D^0\pi^+)\tau^-\bar\nu_\tau$ to analyze the polarization of the $D^{\ast}$ meson in the presence of NP.

\section{Effective operators and helicity amplitudes}
%%%%%%%%%%%%%%%%%%%%%%%%%%%%%%%%%%%%%%%%%%%%%%%%%%%%%%%%%%%%%%%%%%%%%%%%%%%%%%%%%%
Assuming that all neutrinos are left-handed and that NP effects only influence
leptons of the third generation, the effective Hamiltonian for the quark-level
transition $b \to c \tau^- \bar{\nu}_{\tau}$ is given by
\begin{eqnarray*}
{\mathcal H}_{eff} &=&\frac{4G_F}{\sqrt{2}} V_{cb}\left[\mathcal{O}_{V_L}+
\sum\limits_{X=S_L,S_R,V_L,V_R,T_L} X\mathcal{O}_{X}\right],
\end{eqnarray*}
where the four-Fermi operators $\mathcal{O}_{X}$ are defined as $(i=L,R)$
\begin{eqnarray*}
\mathcal{O}_{V_i} &=&
\left(\bar{c}\gamma^{\mu}P_ib\right)
\left(\bar{\tau}\gamma_{\mu}P_L\nu_{\tau}\right),\\
\mathcal{O}_{S_i} &=& \left(\bar{c}P_ib\right)\left(\bar{\tau}P_L\nu_{\tau}\right),
\\
\mathcal{O}_{T_L} &=& \left(\bar{c}\sigma^{\mu\nu}P_Lb\right)
\left(\bar{\tau}\sigma_{\mu\nu}P_L\nu_{\tau}\right).
\end{eqnarray*}
Here, $\sigma_{\mu\nu}=i\left[\gamma_{\mu},\gamma_{\nu}\right]/2$, 
$P_{L,R}=(1\mp\gamma_5)/2$ are the left and right projection operators, and $X$'s are the complex Wilson coefficients governing the NP contributions, which are equal to zero in the SM.

The invariant form factors describing the hadronic transitions $\bar{B}^0 \to D$
and $\bar{B}^0 \to D^\ast$ are defined as follows:
\begin{eqnarray*}
\langle D(p_2)
|\bar{c} \gamma^\mu b
| \bar{B}^0(p_1) \rangle
&=& F_+(q^2) P^\mu + F_-(q^2) q^\mu,\\
\langle D(p_2)
|\bar{c}b
| \bar{B}^0(p_1) \rangle &=& (m_1+m_2)F^S(q^2),\\
\langle D(p_2)|\bar{c}\sigma^{\mu\nu}(1-\gamma^5)b|\bar{B}^0(p_1)\rangle 
&=&\frac{iF^T(q^2)}{m_1+m_2}\left(P^\mu q^\nu - P^\nu q^\mu 
+i \varepsilon^{\mu\nu Pq}\right),\\
\langle D^\ast(p_2)
|\bar{c} \gamma^\mu(1\mp\gamma^5)b
| \bar{B}^0(p_1) \rangle
&=& \frac{\epsilon^{\dagger}_{2\alpha}}{m_1+m_2}
\Big[ \mp g^{\mu\alpha}PqA_0(q^2) \pm P^{\mu}P^{\alpha}A_+(q^2)\\
&&\pm q^{\mu}P^\alpha A_-(q^2) 
+ i\varepsilon^{\mu\alpha P q}V(q^2)\Big],
\\
\langle D^\ast(p_2)
|\bar{c}\gamma^5 b
| \bar{B}^0(p_1) \rangle &=& \epsilon^\dagger_{2\alpha}P^\alpha G^S(q^2),
\\
\langle D^\ast(p_2)|\bar{c}\sigma^{\mu\nu}(1-\gamma^5)b|\bar{B}^0(p_1)\rangle
&=&-i\epsilon^\dagger_{2\alpha}\Big[
\left(P^\mu g^{\nu\alpha} - P^\nu g^{\mu\alpha} 
+i \varepsilon^{P\mu\nu\alpha}\right)G_1^T(q^2)\\
&&+\left(q^\mu g^{\nu\alpha} - q^\nu g^{\mu\alpha}
+i \varepsilon^{q\mu\nu\alpha}\right)G_2^T(q^2)\\
&&+\left(P^\mu q^\nu - P^\nu q^\mu 
+ i\varepsilon^{Pq\mu\nu}\right)P^\alpha\frac{G_0^T(q^2)}{(m_1+m_2)^2}
\Big],
\end{eqnarray*}
where $P=p_1+p_2$, $q=p_1-p_2$, and $\epsilon_2$ is the polarization vector
of the $D^\ast$ meson which satisfies the condition $\epsilon_2^\dagger\cdot p_2=0$.
The particles are on their mass shells: $p_1^2=m_1^2=m_{\bar{B}^0}^2$ and
 $p_2^2=m_2^2=m_{D^{(\ast)}}^2$.
%%%

Using the helicity technique first described 
in~\cite{Korner-Schuler} and
further discussed in our recent papers~\cite{Ivanov:2015tru,Ivanov:2016qtw}, one obtains
the ratio of branching fractions $R_{D^{(*)}}(q^2)$ as follows:
\begin{equation*}
R_{D^{(*)}}(q^2)
=\left(\frac{q^2-m_\tau^2}{q^2-m_\mu^2}\right)^2\frac{{\cal H}_{tot}^{D^{(\ast)}}}{\sum\limits_{n}|H_{n}|^2+\delta_\mu \big(\sum\limits_{n}|H_{n}|^2+3|H_{t}|^2\big)},
\end{equation*}
%%%
\begin{eqnarray*}
{\cal H}_{tot}^{D}
&=&
|1+g_V|^2\left[|H_0|^2+\delta_\tau(|H_0|^2+3|H_t|^2) \right]+\frac{3}{2}|g_S|^2 |H_P^S|^2\\
&&+ 3\sqrt{2\delta_\tau} {\rm Re}g_S H_P^S H_t
+8|T_L|^2 ( 1+4\delta_\tau) |H_T|^2
+12\sqrt{2\delta_\tau} {\rm Re}T_L H_0 H_T,\\
{\cal H}_{tot}^{D^\ast}
&=&(|1+V_L|^2+|V_R|^2)\Big[\sum\limits_{n}|H_{n}|^2+\delta_\tau \Big(\sum\limits_{n}|H_{n}|^2+3|H_{t}|^2\Big) \Big]+\frac{3}{2}|g_P|^2|H^S_V|^2\\
&&-2 {\rm Re}V_R\big[(1+\delta_\tau) (|H_{0}|^2+2H_{+}H_{-})+3\delta_\tau |H_{t}|^2 \big]
-3\sqrt{2\delta_\tau} {\rm Re}g_P H^S_V H_{t}\\
&&+8|T_L|^2 (1+4\delta_\tau)\sum\limits_{n}|H_T^n|^2
-12\sqrt{2\delta_\tau} {\rm Re}T_L\sum\limits_{n} H_{n}H_T^n.
\end{eqnarray*}
Here, $\delta_\ell = m^2_\ell/2q^2$ is the helicity flip factor, $g_V\equiv V_L+V_R$, $g_S\equiv S_L+S_R$, $g_P\equiv S_L-S_R$, and the index $n$ runs through $(0,+,-)$. The definition of the hadronic helicity amplitudes $H$ in terms of the invariant form factors is presented in the Appendix of~\cite{Ivanov:2017mrj}. Note that we do not consider interference terms between different NP operators since we assume the dominance of only one NP operator besides the SM contribution. 
%---------------------------------------------------------------------------
\section{Form factors in the CCQM}
As has been discussed in detail in~\cite{Ivanov:2016qtw} we calculate the
      current-induced $B \to D^{(\ast)}$ transitions from their one-loop quark
      diagrams. As a result
the various form factors in our model are represented by threefold integrals
which are calculated by using \textsc{fortran} codes in the full kinematical
momentum 
transfer region $0\le q^2 \le q^2_{max}=(m_{\bar{B}^0}-m_{D^{(\ast)}})^2$. Our numerical results for the form factors are well represented
by a double-pole parametrization
\begin{equation*}
F(q^2)=\frac{F(0)}{1 - a s + b s^2}, \quad s=\frac{q^2}{m_1^2}. 
\end{equation*}
The parameters of the  form factors for the $\bar{B}^0 \to D$ and $\bar{B}^0 \to D^\ast$ transitions are listed in Table~\ref{tab:ff}.
\begin{table}[htbp] 
\setlength{\tabcolsep}{2pt}
\begin{tabular}{|l|cccccccc|cccc|}
\hline
\multicolumn{1}{|c|}{} &\multicolumn{8}{c|}{$\bar{B}^0 \to D^\ast$} 
                      &\multicolumn{4}{c|}{$\bar{B}^0 \to D$} \\
 & $ A_0 $ & $  A_+  $ & $  A_-  $ & $  V  $ 
 & $ G^S $ & $G_0^T$ & $G_1^T$ & $  G_2^T$  & $F_+$ & $F_-$ & $F^S$ & $F^T$ 
 \\
\hline
$F(0)$ &  1.62 & 0.67  & $-0.77$ & 0.77 & $-0.50$ & $-0.073$ & 0.73 & $-0.37$ &  0.79   & $-0.36$ &  0.80 & 0.77  
\\
$a$    &  0.34 & 0.87  &  0.89 & 0.90 & 0.87 & 1.23 & 0.90 & 0.88 &  0.75   &  0.77 &  0.22 & 0.76  
\\
$b$    & $-0.16$ & 0.057 & 0.070 & 0.075 & 0.060 & 0.33 & 0.074 & 0.065 &  0.039  & 0.046 & $-0.098$ & 0.043 
\\ 
$F(q^2_{\rm max})$ &  1.91 & 0.99  & $-1.15$ & 1.15 & $-0.74$ & $-0.13$ & 1.10 & $-0.55$ &  1.14   & $-0.53$ &  0.89 & 1.11  
\\
$F^{HQL}(q^2_{\rm max})$ &  1.99 & 1.12  & $-1.12$ & 1.12 & $-0.62$ & 0 & 1.12 & $-0.50$ &  1.14   & $-0.54$ &  0.88 & 1.14  
\\
\hline
\end{tabular}
\caption{Parameters of the dipole approximation for  $\bar{B}^0 \to D^{(\ast)}$ form factors. Zero-recoil values of the form factors are also listed for comparison with the HQET.}
\label{tab:ff}
\end{table}
We also list the zero-recoil values of the form factors for comparison with the corresponding HQET results which can e.g. be found in~\cite{Ivanov:2016qtw}. The agreement between the two sets of
zero-recoil values is within $10 \%$. It is worth mentioning that we obtain a nonzero result for the form factor
$G_0^T$ at zero recoil, which is predicted to vanish in the HQET. 

We note that in~\cite{Ivanov:2015tru} the HQL in our approach was explored in great detail  for the heavy-to-heavy $\bar{B}^0\to D^{(\ast)}$ transitions. In~\cite{Ivanov:2015tru} we also calculated the Isgur-Wise function and considered the near-recoil behavior of the form factors. A brief discussion of the subleading corrections to the HQL arising from finite quark masses can be found in Appendix~B of~\cite{Ivanov:2016qtw}. 
Finally, we briefly discuss some error estimates within our model.
We fix our model parameters (the constituent quark masses, the infrared cutoff,
and the hadron size parameters) by minimizing the functional
$\chi^2 = \sum\limits_i\frac{(y_i^{\rm expt}-y_i^{\rm theor})^2}{\sigma^2_i}$
where $\sigma_i$ is the experimental  uncertainty.
If $\sigma$ is too small then we take its value of 10$\%$.
Moreover, we observed that the errors of the fitted parameters 
are of the order of  10$\%$.
Thus we estimate the model uncertainties to lie within 10$\%$.
%---------------------------------------------------------------------------
\section{Experimental constraints}
%%%%%
%%%%%%%%%%%
\begin{wrapfigure}{r}{0.64\textwidth}
\begin{tabular}{rr}
\includegraphics[scale=0.34]{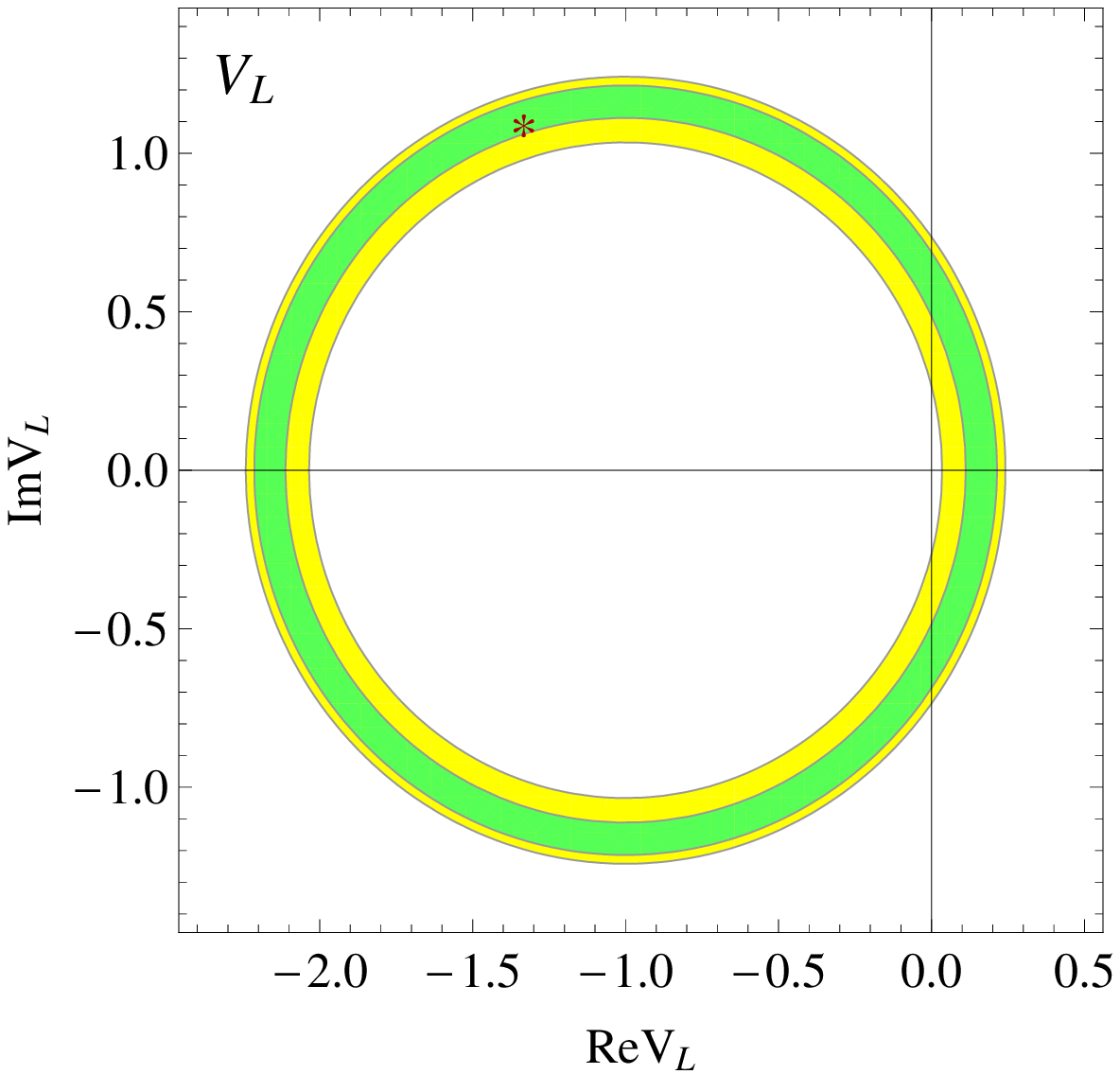}&
\includegraphics[scale=0.34]{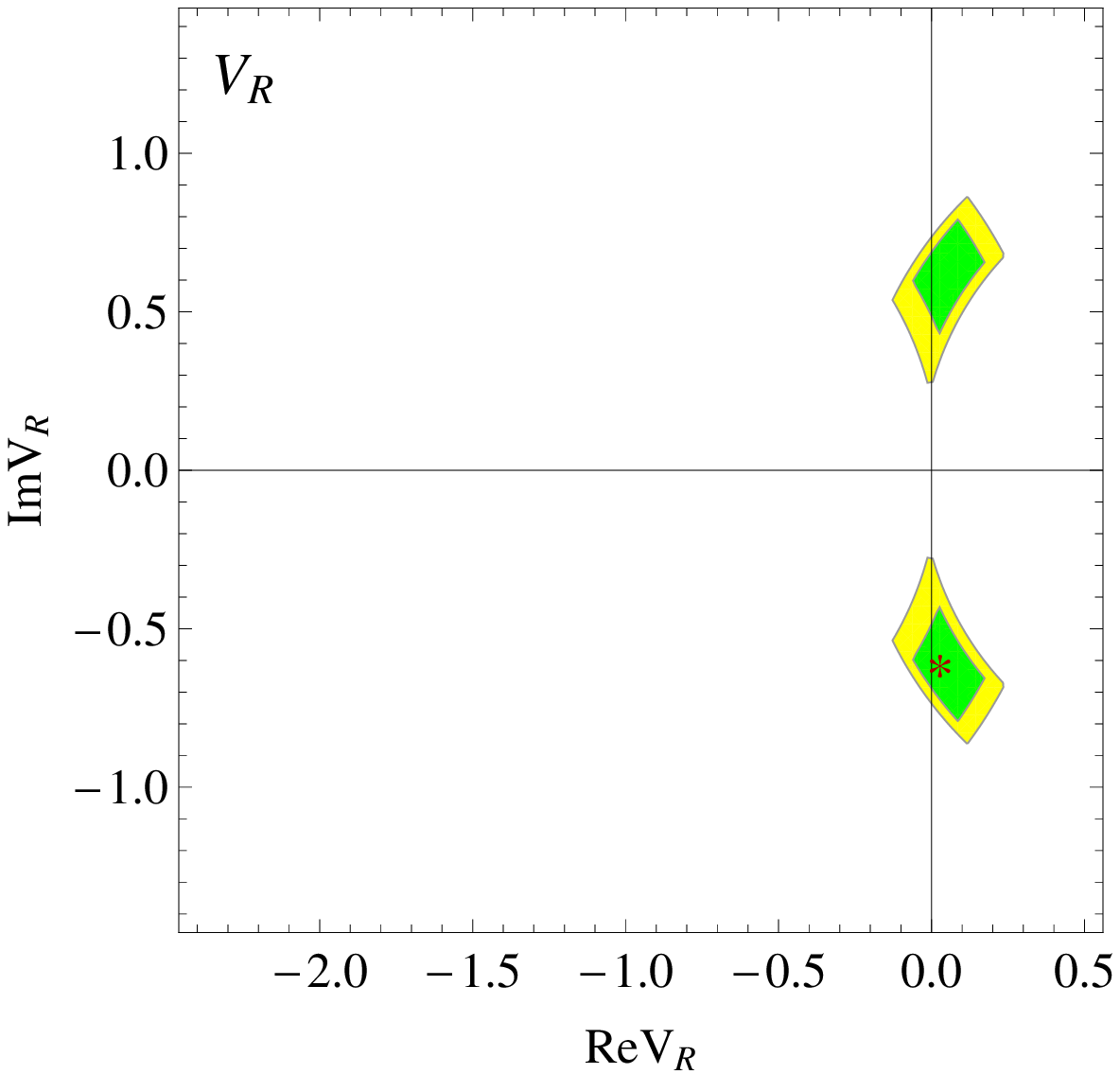}\\
\includegraphics[scale=0.34]{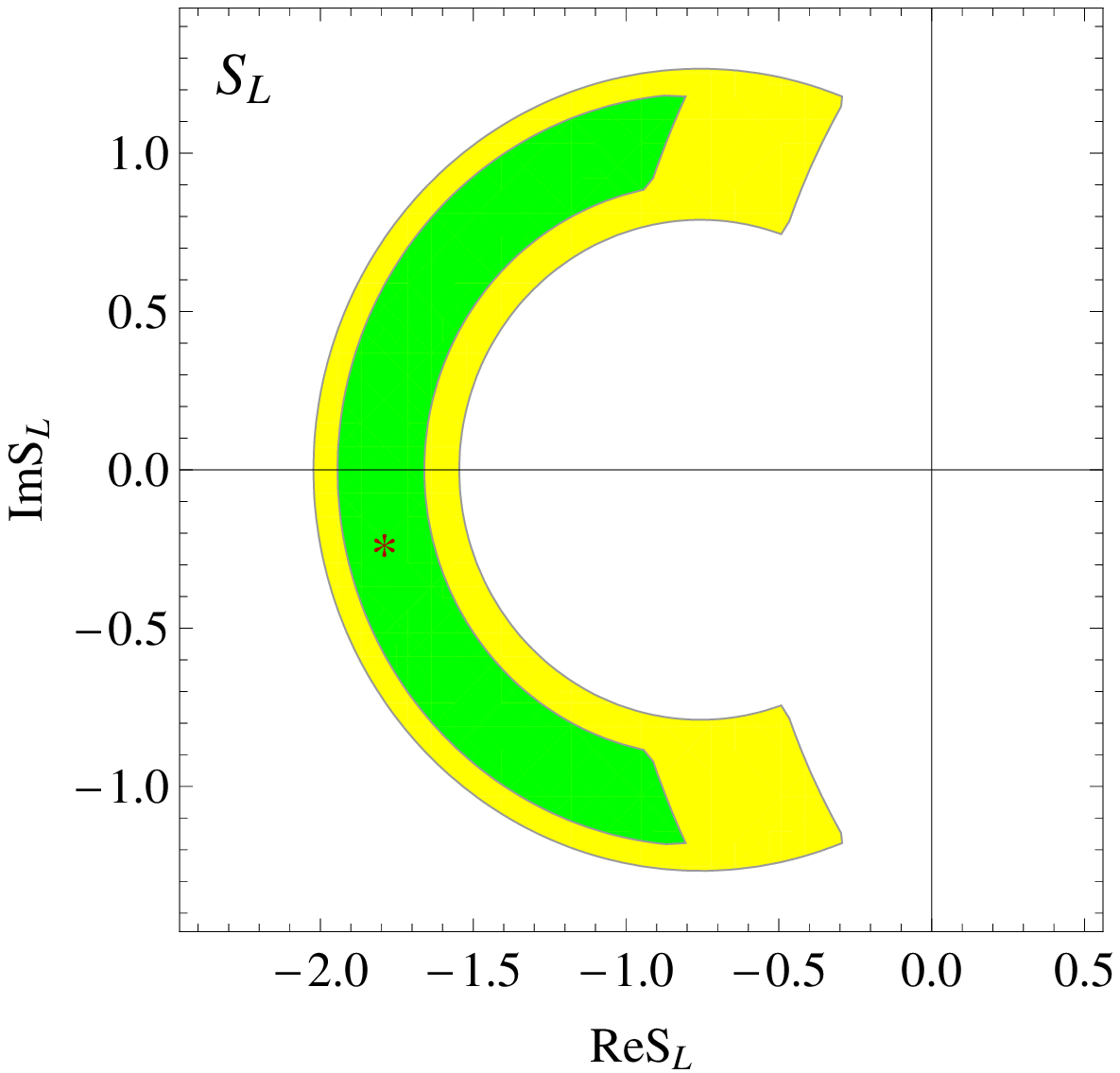}
& 
\includegraphics[scale=0.34]{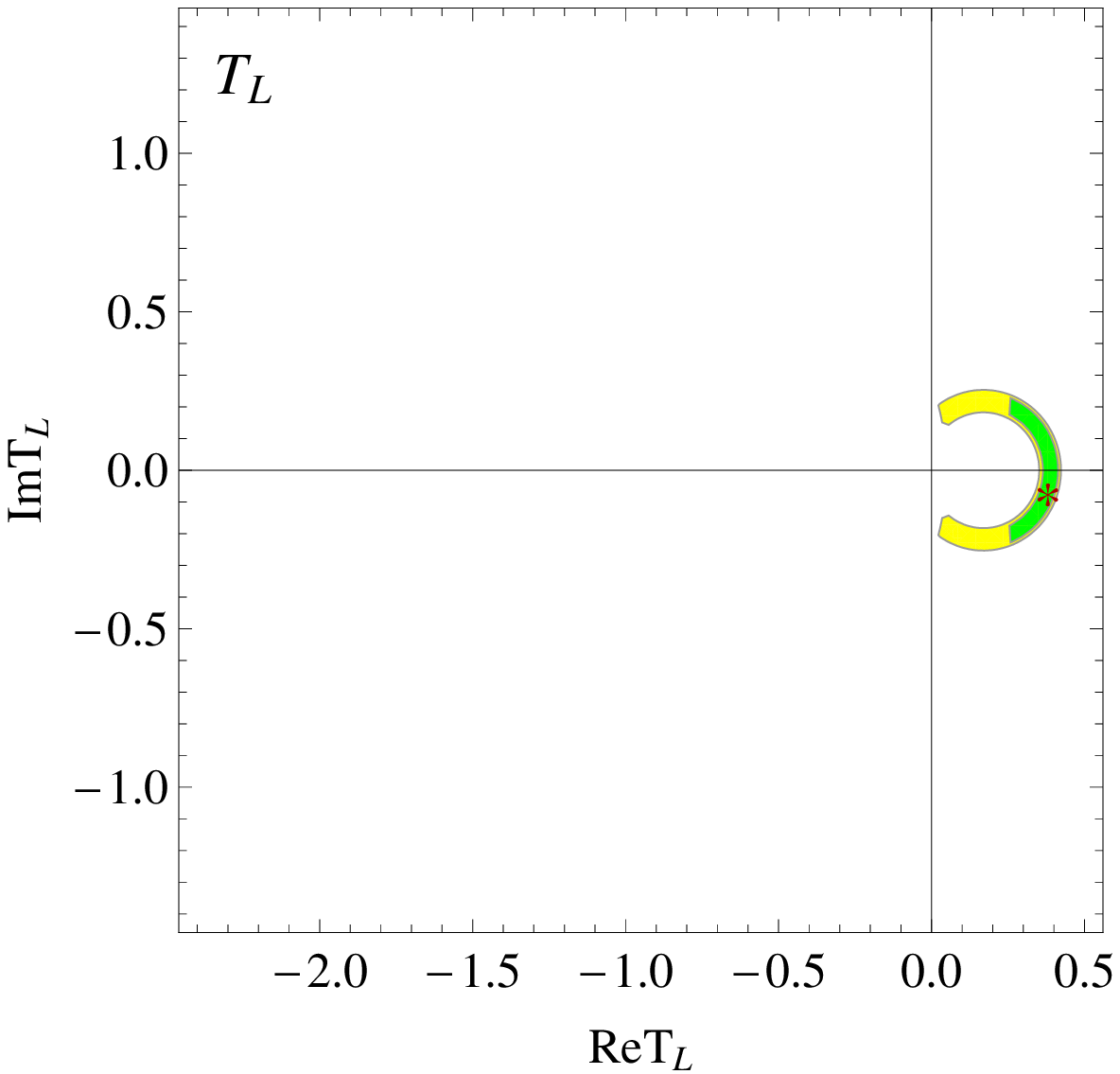}
\end{tabular}
\caption{Constraints on the Wilson coefficients $V_L$, $V_R$, $S_L$, and $T_L$ within $1\sigma$ (green, dark) and $2\sigma$ (yellow, light). No value of $S_R$ is allowed within $2\sigma$. The best-fit value in each case is denoted with the symbol $\ast$.}
\label{fig:constraint}
\end{wrapfigure}
%%%
Within the SM (without any NP operators) our model calculation yields
$R(D)=0.267$ and $R(D^\ast)=0.238$,
which are consistent with other SM predictions given in~\cite{Na:2015kha, Fajfer:2012vx} within $10\%$.

 Assuming the dominance of only one NP operator at a time (besides the SM one), we compare the calculated ratios $R_{D^{(\ast)}}$ with the current experimental data $R_D = 0.406 \pm 0.050$ and $R_{D^\ast} = 0.311\pm 0.016$  and obtain the allowed regions for the NP couplings as shown in Fig.~\ref{fig:constraint}. 

It is important to note that while determining these regions, we also take into account a theoretical error of $10\%$ for the ratios $R(D^{(\ast)})$. The vector operators $\mathcal{O}_{V_{L,R}}$ and the left scalar operator $\mathcal{O}_{S_L}$ are favored while there is no allowed region for the right scalar operator $\mathcal{O}_{S_R}$ within $2\sigma$. Therefore we will not consider $\mathcal{O}_{S_R}$ in what follows. The tensor operator $\mathcal{O}_{T_L}$ is less favored, but it can still well explain the current experimental results. In each allowed region at $2\sigma$ we find the best-fit value for each NP coupling. The best-fit couplings read $V_L =-1.33+i1.11$, $V_R =0.03-i0.60$, $S_L =-1.79-i0.22$, $T_L =0.38-i0.06$, and are marked with an asterisk.
%%%%%%%%%%%%%%%%%%%%%%%%%%%%%%%%%%%%%%%%%%
\section{The cascade decay 
$\bar{B}^0\to D^{\ast+}(\to D^0\pi^+)\tau^-\bar{\nu}_\tau$ and 
the\\ angular observables}
%%%%%%%%%%
\begin{wrapfigure}{r}{0.55\textwidth}
\includegraphics[scale=0.35]{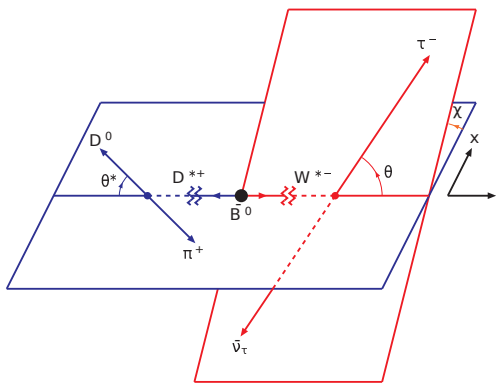}
\caption{Definition of the angles $\theta$, $\theta^\ast$ and $\chi$ in
the cascade decay $\bar B^0\to D^{\ast+}(\to D^0\pi^+)\tau^-\bar\nu_\tau$.}
\label{fig:bdangl}
\end{wrapfigure}
%%%%%
\subsection{The fourfold distribution}
\begin{wrapfigure}{r}{0.61\textwidth}
\begin{tabular}{rr}
\includegraphics[scale=0.3]{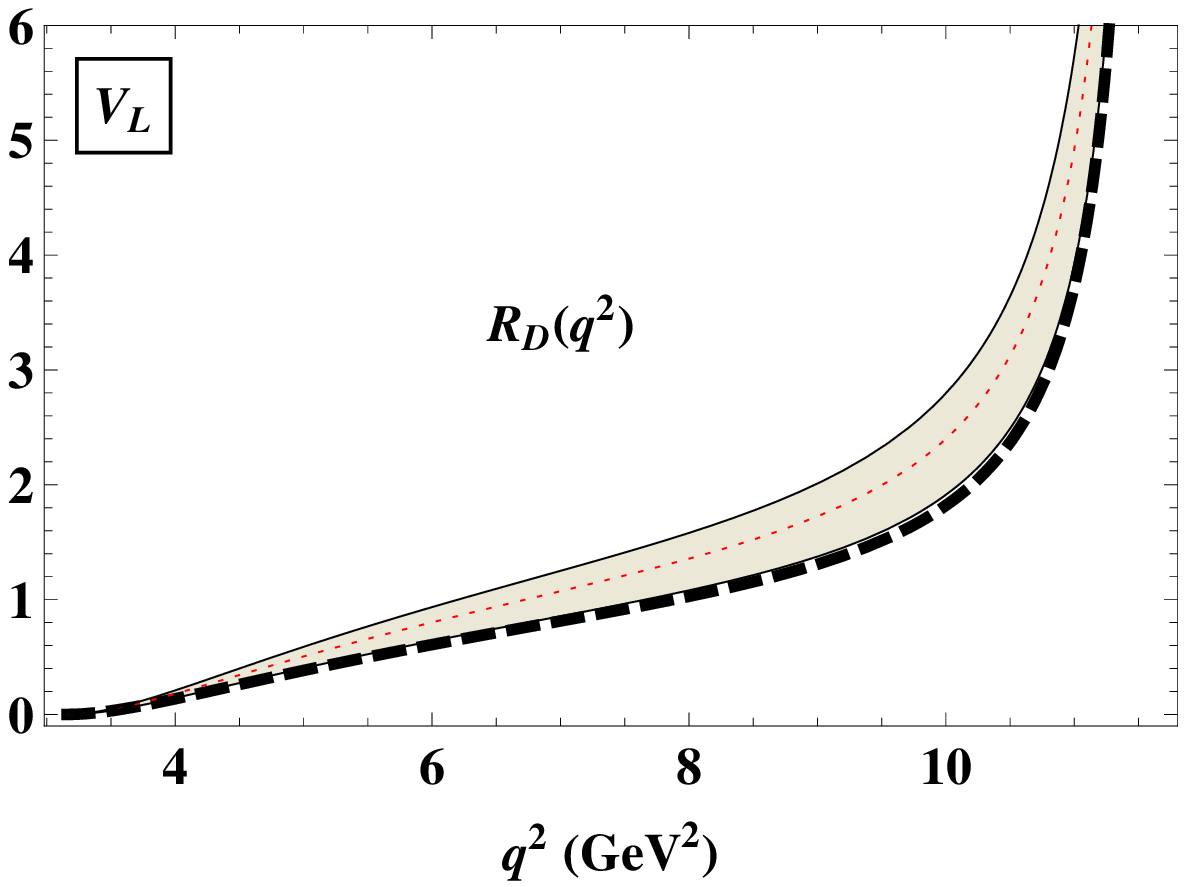}&
\includegraphics[scale=0.3]{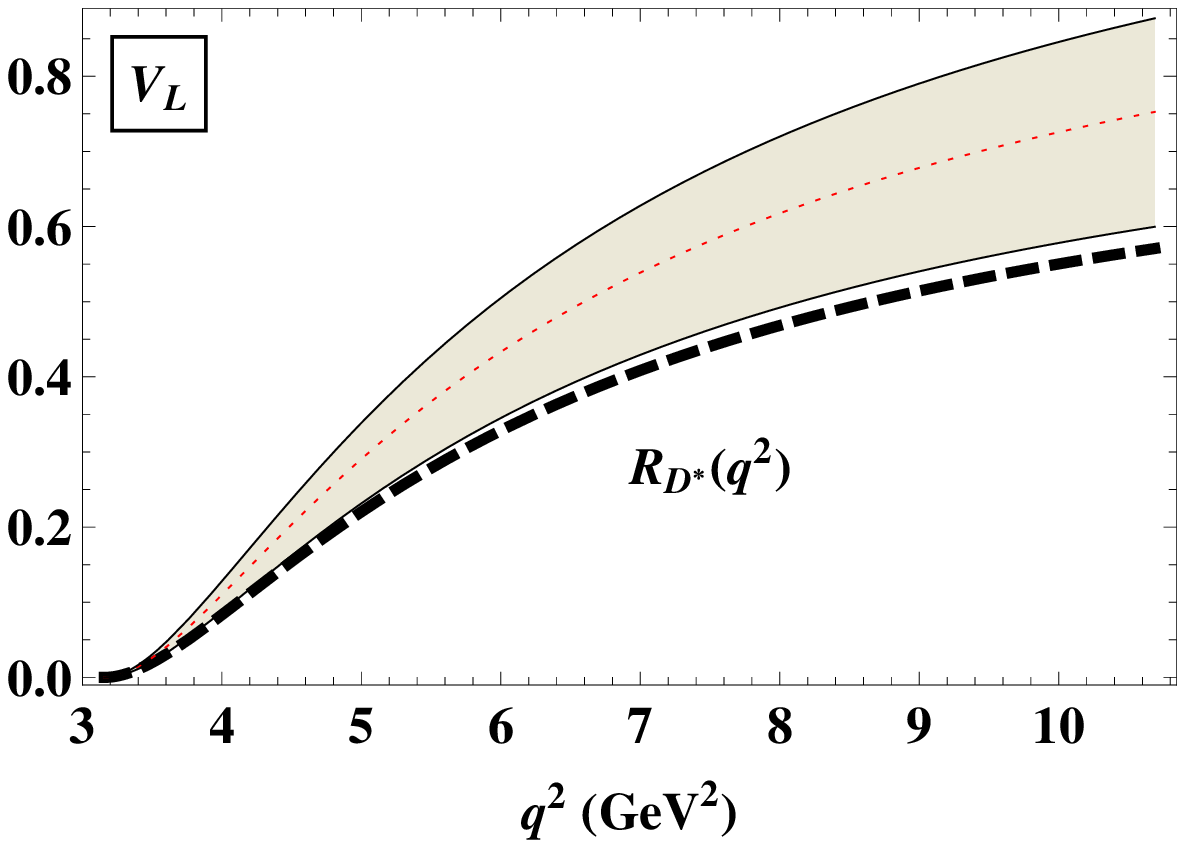}\\
\includegraphics[scale=0.3]{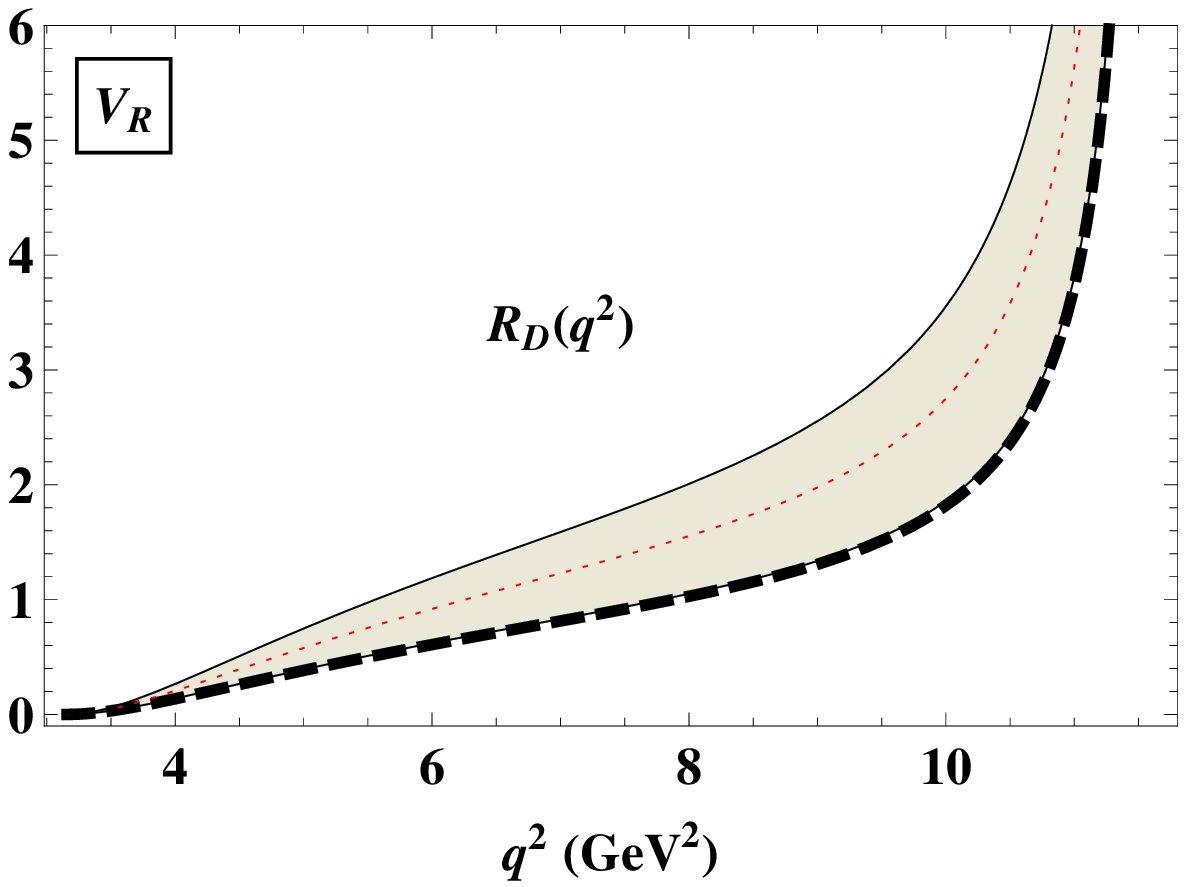}&
\includegraphics[scale=0.3]{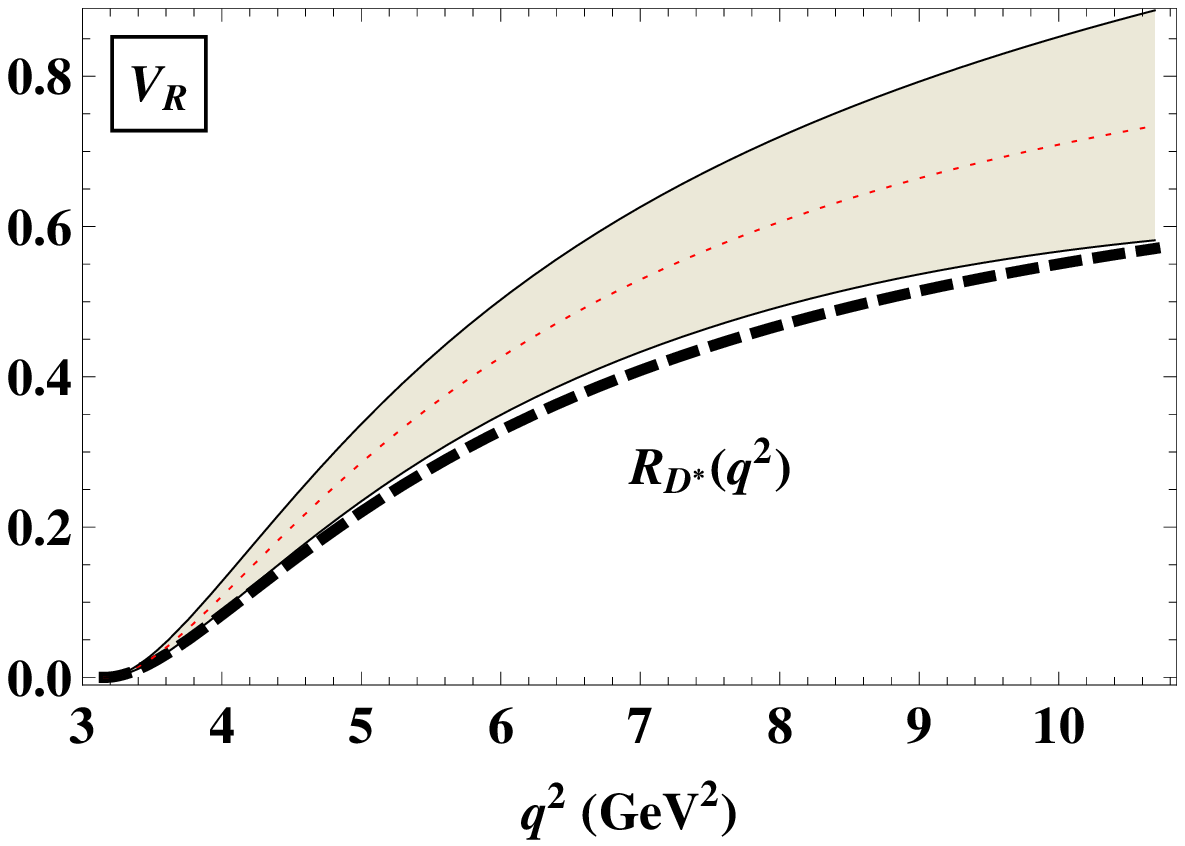}
\\
\includegraphics[scale=0.3]{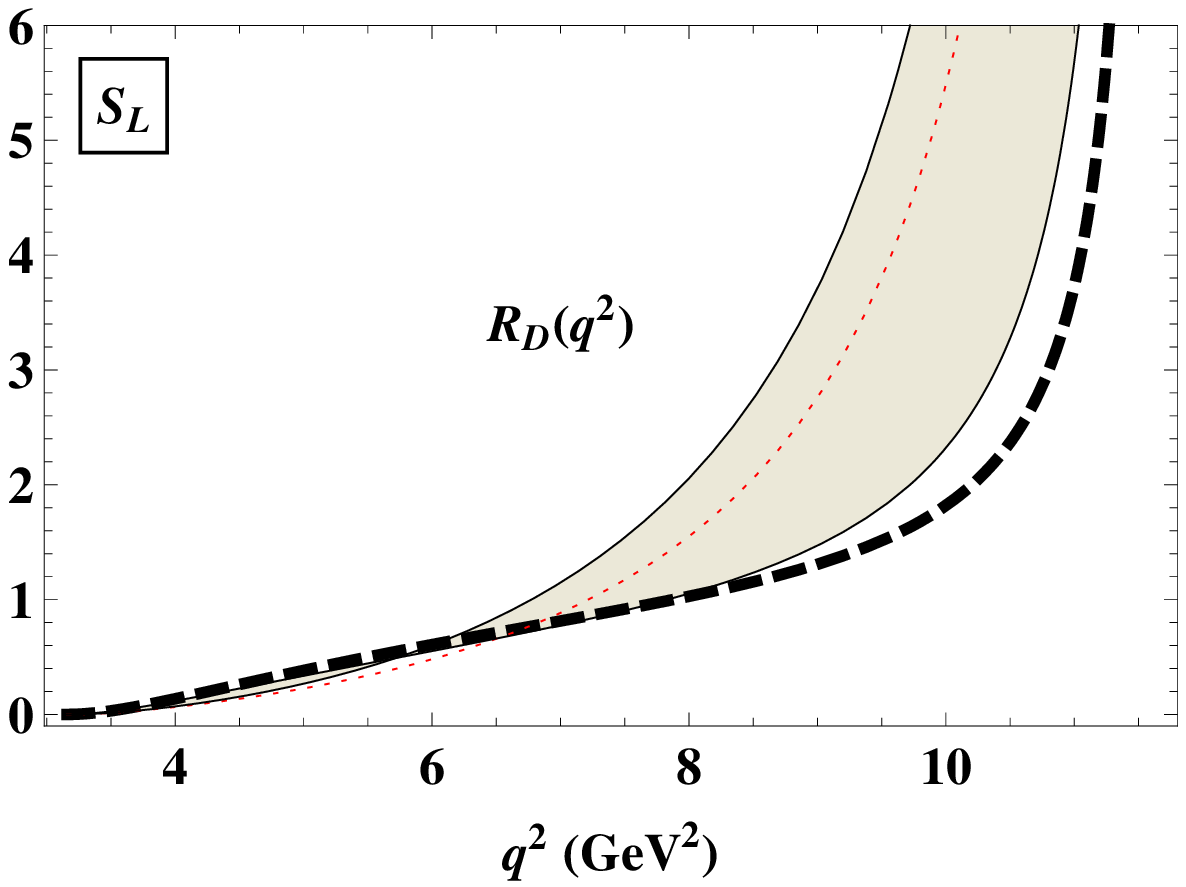}&
\includegraphics[scale=0.3]{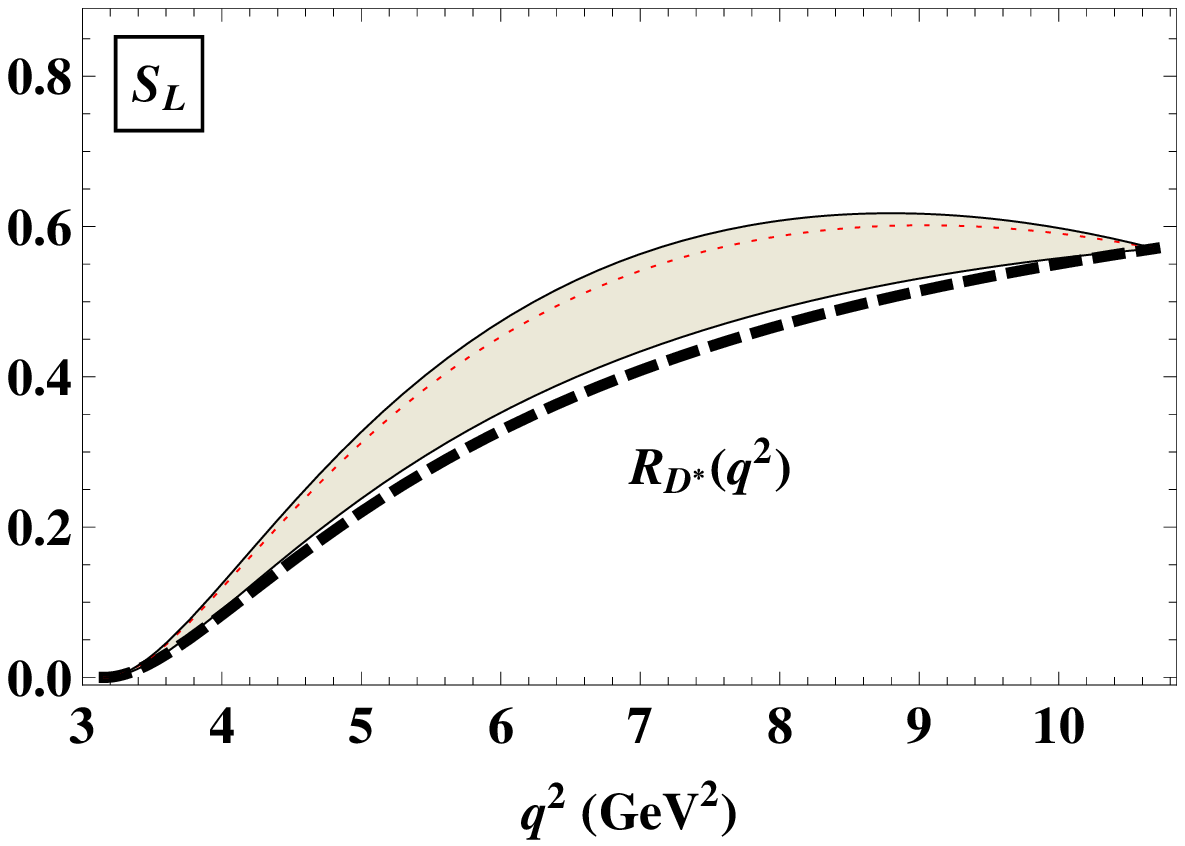}\\
\includegraphics[scale=0.3]{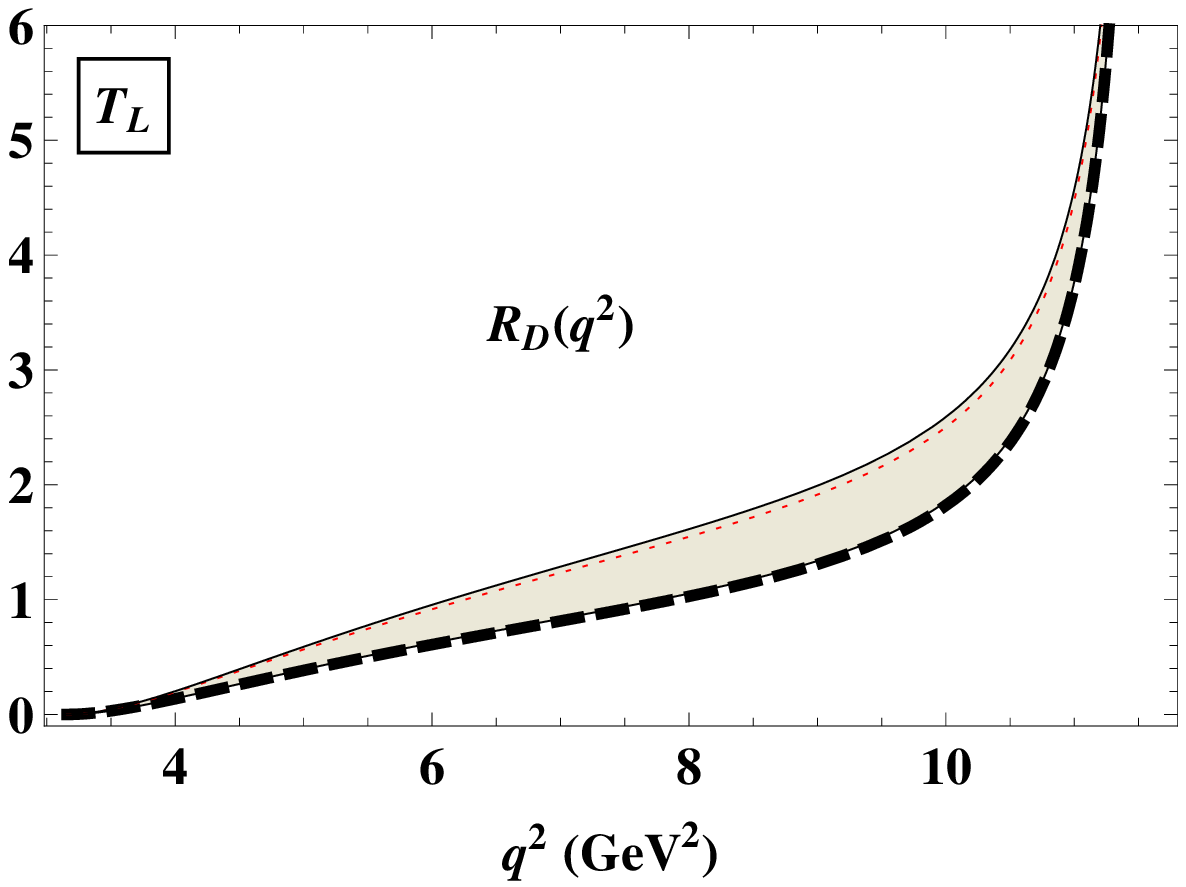}&
\includegraphics[scale=0.3]{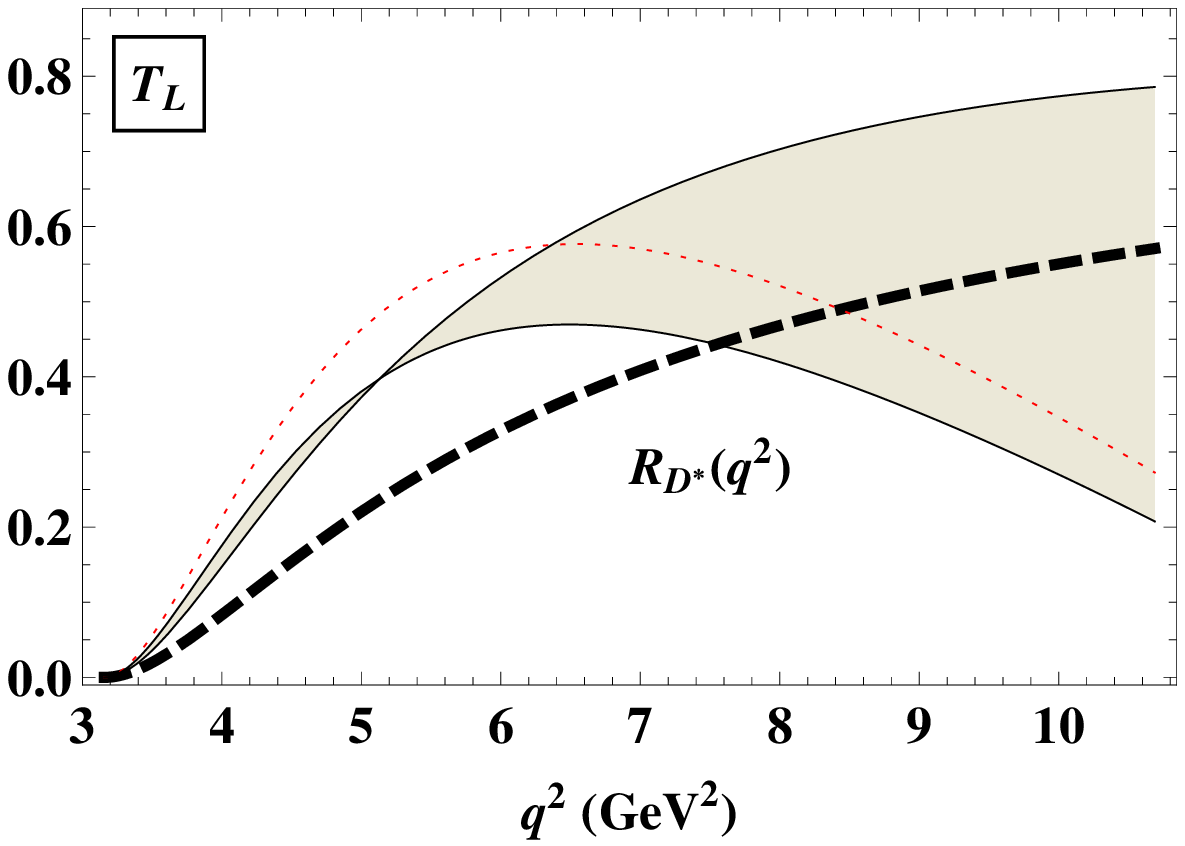}
\end{tabular}
\caption{$R_{D}(q^2)$ (left) and $R_{D^\ast}(q^2)$ (right). The thick black dashed lines are the SM prediction; the gray bands include NP effects corresponding to the $2\sigma$ allowed regions in Fig.~\ref{fig:constraint}; the red dotted lines represent the best-fit values.}
\label{fig:RD}
\end{wrapfigure}

In order to analyze NP effects on the polarization of the $D^\ast$ meson one uses the cascade decay $\bar{B}^0\to D^{\ast+}(\to D^0\pi^+)\tau^-\bar{\nu}_\tau$. A detailed derivation of the fourfold angular distribution (without NP) can be found in our paper~\cite{Faessler:2002ut}. The three angles $\theta$, $\theta^\ast$, and $\chi$ in the distribution are defined in Fig.~\ref{fig:bdangl}
One has
\begin{eqnarray*}
\frac{d^4\Gamma(\bar{B}^0\to D^{\ast+}(\to D^0\pi^+) \tau^-\bar\nu_\tau)}
     {dq^2 d\cos\theta d\chi d\cos\theta^\ast}\\ 
=\frac{9}{8\pi}|N|^2J(\theta,\theta^\ast,\chi),
\end{eqnarray*}
where
\begin{equation*}
|N|^2=
\frac{G_F^2 |V_{cb}|^2 |{\bf p_2}| q^2 v^2}{(2\pi)^3 12 m_1^2}\mathcal{B}(D^\ast\to D\pi).
\end{equation*}
The full angular distribution $J(\theta,\theta^\ast,\chi)$ is written as
\begin{eqnarray*}
\lefteqn{J(\theta,\theta^\ast,\chi)}\\
&=& J_{1s}\sin^2\theta^\ast + J_{1c}\cos^2\theta^\ast
+(J_{2s}\sin^2\theta^\ast + J_{2c}\cos^2\theta^\ast)\cos2\theta\\
&&+J_3\sin^2\theta^\ast \sin^2\theta \cos2\chi
+J_4\sin2\theta^\ast \sin2\theta \cos\chi\\
&&+J_5\sin2\theta^\ast \sin\theta \cos\chi
+(J_{6s}\sin^2\theta^\ast+J_{6c}\cos^2\theta^\ast)\cos\theta\\
&&+J_7\sin2\theta^\ast \sin\theta \sin\chi
+J_8\sin2\theta^\ast \sin2\theta \sin\chi
+J_9\sin^2\theta^\ast \sin^2\theta \sin2\chi ,
\end{eqnarray*}
where $J_{i(a)}$ $(i=1,\dots,9; a=s,c)$ are the angular observables. Their explicit expressions in terms of helicity amplitudes and Wilson coefficients can be found in our paper~\cite{Ivanov:2016qtw}.
 The fourfold distribution allows one to define a large set of observables which can help probe NP in the decay. 
First, by integrating the angular decay distribution over all angles one obtains 
\begin{equation*}
\frac{d\Gamma(\bar{B}^0\to D^{\ast} \tau^-\bar\nu_\tau)}{dq^2} =
|N|^2 J_{\rm tot} = |N|^2 (J_L+J_T),
\end{equation*}
where $J_L=3J_{1c}-J_{2c}$ and $J_T=2(3J_{1s}-J_{2s})$ are the longitudinal and transverse polarization amplitudes of the $D^\ast$ meson.
In Fig.~\ref{fig:RD} we present the 
$q^{2}$ dependence
of the rate ratios $R_{D^{(\ast)}}(q^{2})$
in different NP scenarios. It is interesting to note that unlike the vector and scalar operators, which tend to increase both ratios, the tensor operator can lead to a decrease of the ratio $R(D^\ast)$ for $q^2 \gtrsim 8~\text{GeV}^2$. Moreover, while the ratio $R(D^\ast)$ is minimally sensitive to the scalar coupling $S_L$ (in comparison with other couplings, i.e. $V_{L,R}$, $T_L$), the ratio $R(D)$ shows maximal sensitivity to $S_L$. These behaviors can help discriminate between different NP operators.
%
%----------------------------------------------------------------------------
%
\subsection{The $\cos\theta$ distribution, the forward-backward asymmetry,  and the lepton-side convexity parameter}
\begin{figure}[htbp]
\begin{tabular}{ccc}
\includegraphics[scale=0.3]{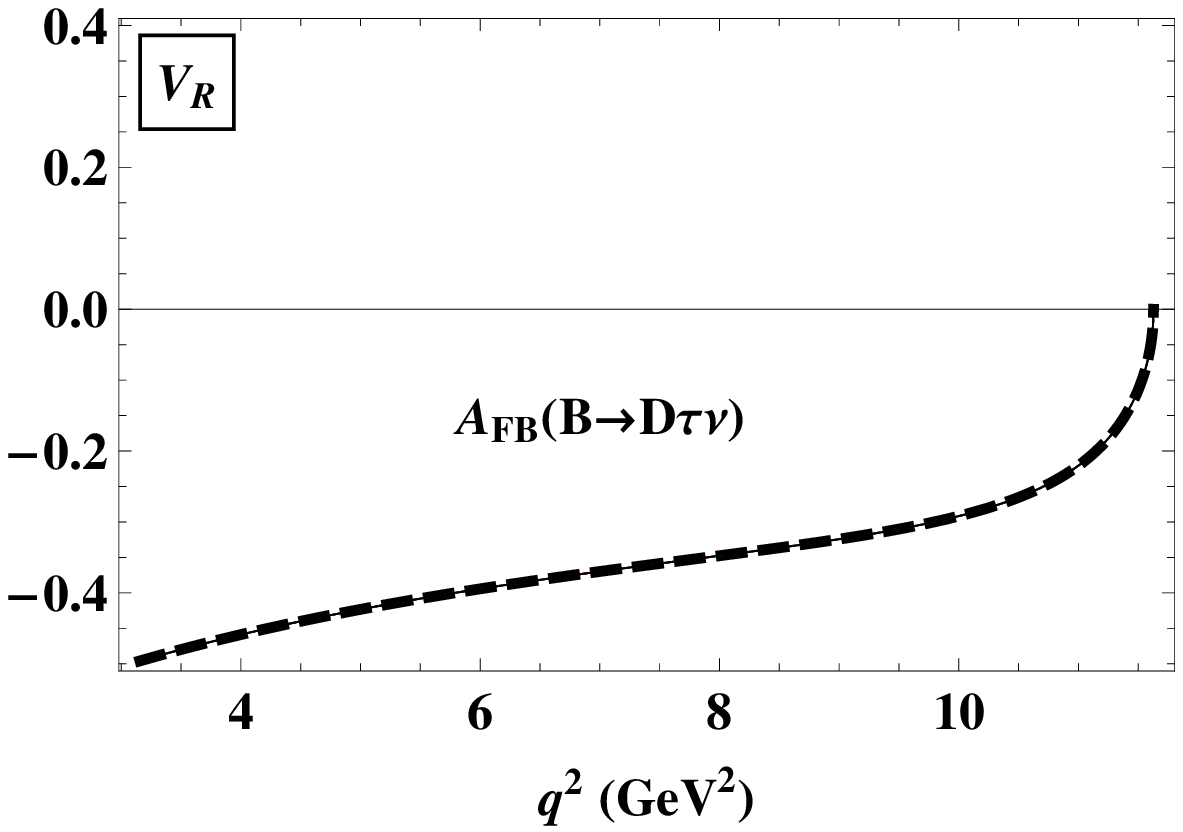}&
\includegraphics[scale=0.3]{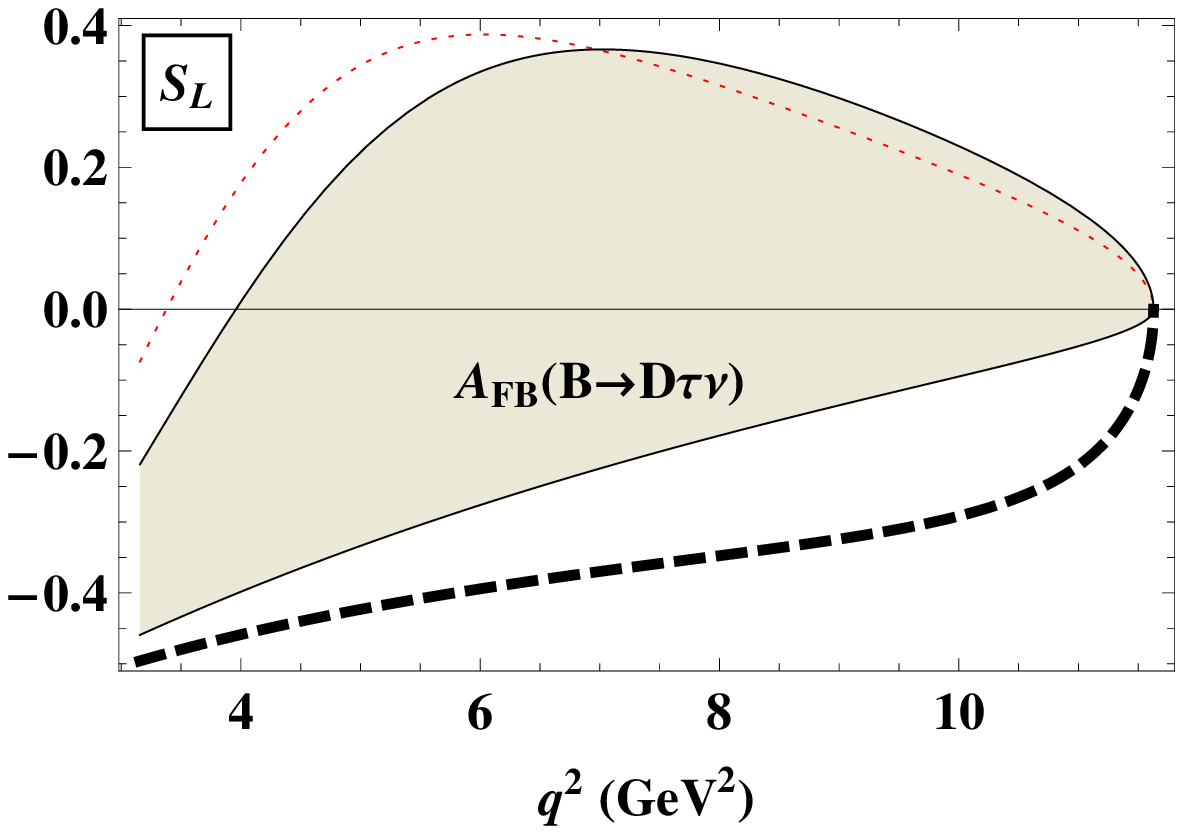}&
\includegraphics[scale=0.3]{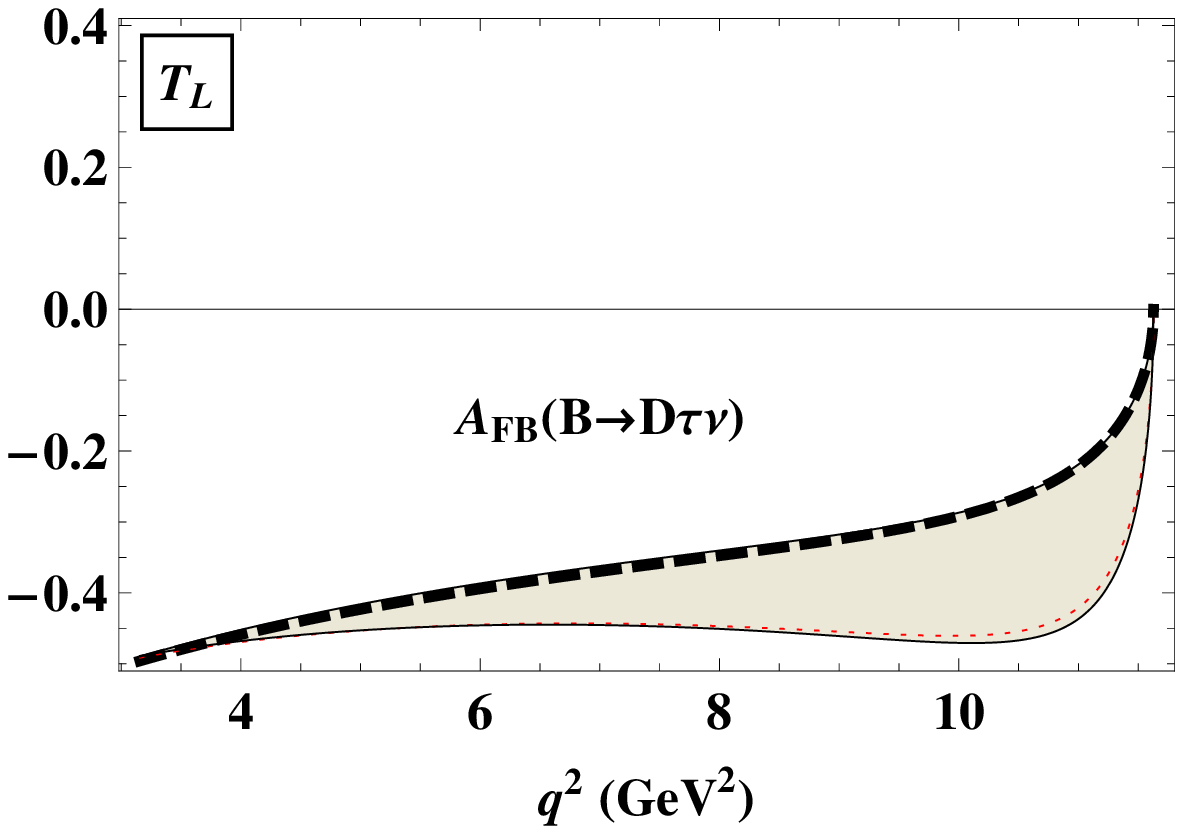}\\
\includegraphics[scale=0.3]{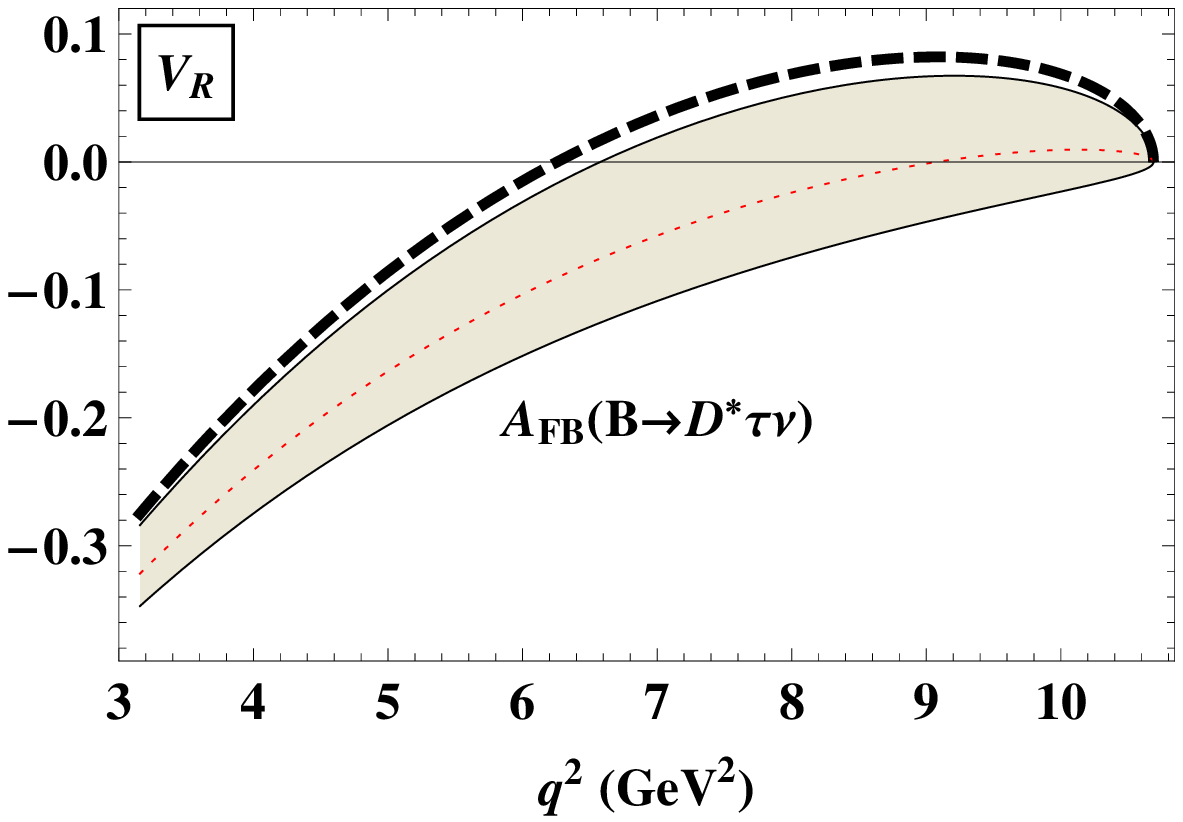}&
\includegraphics[scale=0.3]{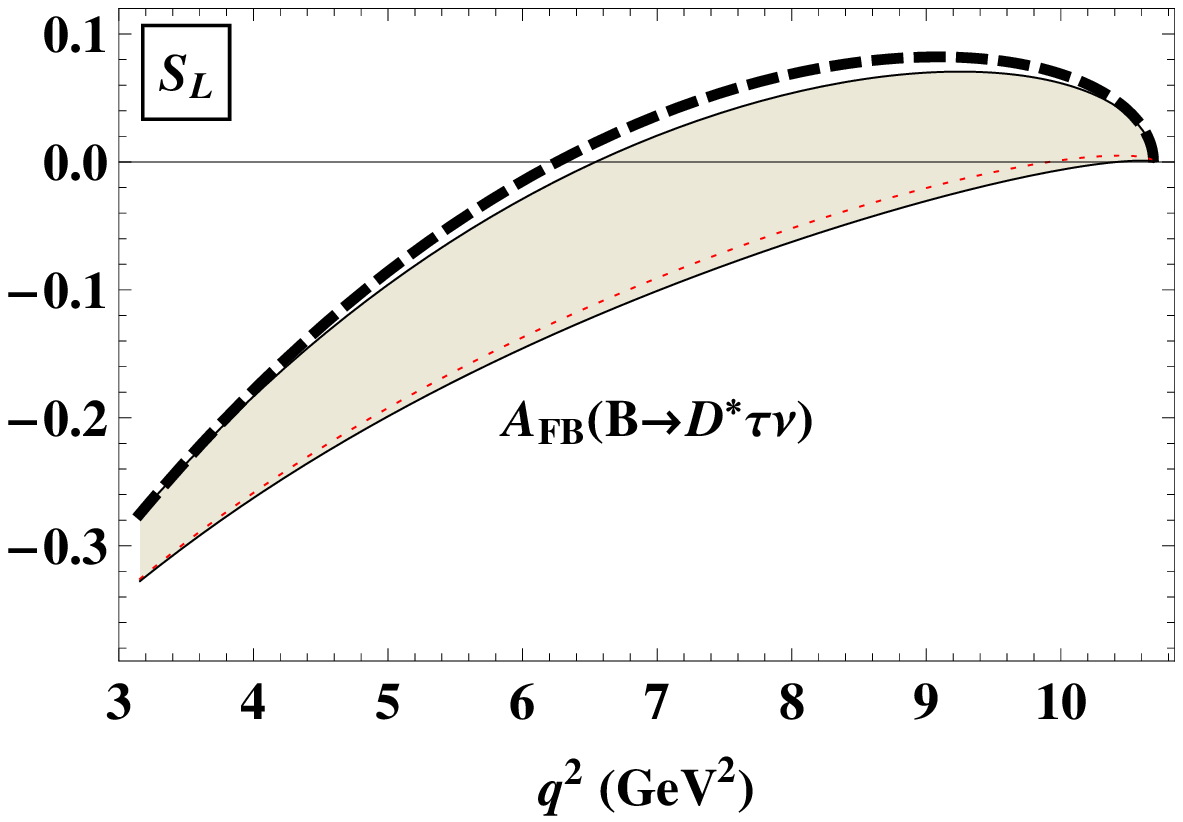}&
\includegraphics[scale=0.3]{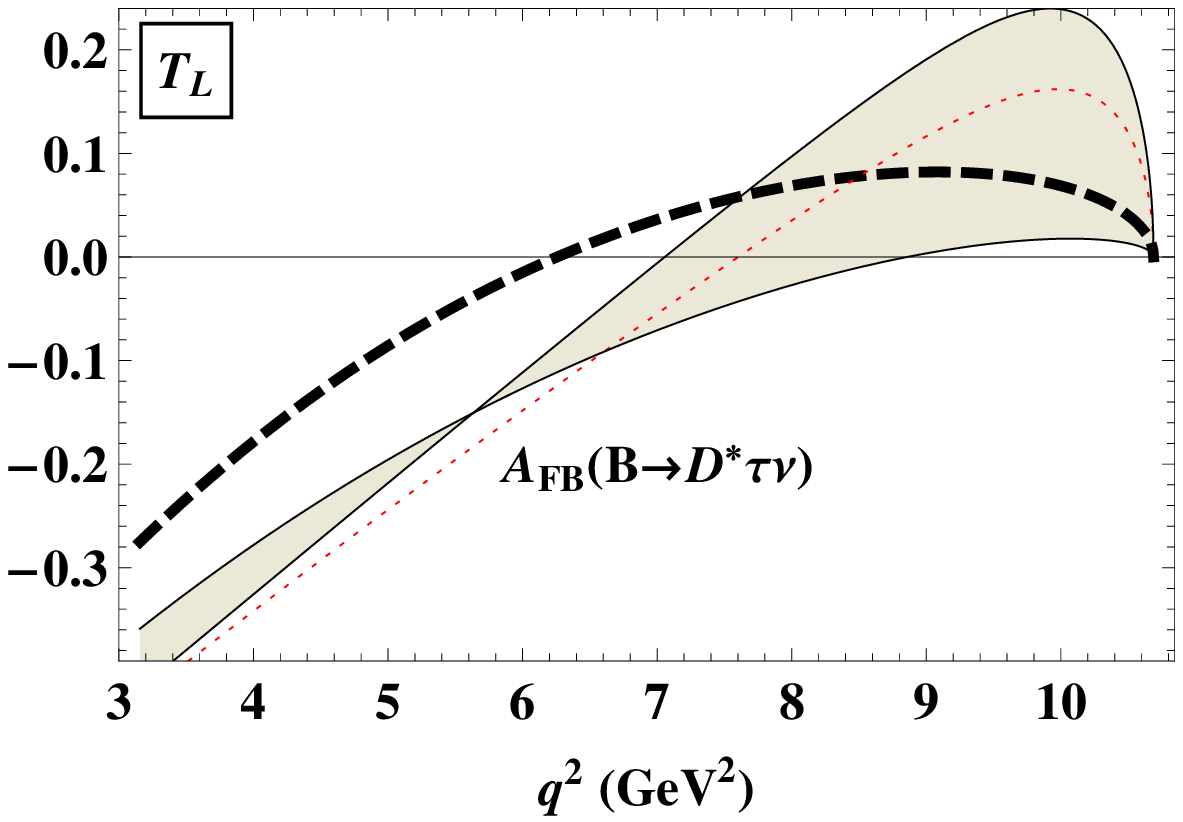}
\end{tabular}
\caption{Forward-backward asymmetry $\mathcal{A}_{FB}$ for $\bar{B}^0 \to D\tau^-\bar\nu_{\tau}$ (upper) and $\bar{B}^0 \to D^\ast\tau^-\bar\nu_{\tau}$ (lower). Notations are the same as in Fig.~\ref{fig:RD}.}
\label{fig:AFB}
\end{figure}

We define a normalized angular decay distribution 
$\widetilde J(\theta^\ast,\theta,\chi)$ through
\begin{equation*}
\widetilde J(\theta^\ast,\theta,\chi)=\frac{9}{8\pi}\frac{J(\theta^\ast,\theta,\chi)}
{J_{\rm tot}},
\end{equation*}
where $J_{\rm tot}=3J_{1c}+6J_{1s}-J_{2c}-2J_{2s}$.
The normalized angular decay distribution 
$\widetilde J(\theta^\ast,\theta,\chi)$ obviously integrates to $1$ after
$\cos\theta^\ast,\,\cos\theta$, and $\chi$ integration. By integrating 
the fourfold distribution over $\cos\theta^\ast$ and 
$\chi$ one obtains the differential $\cos\theta$ distribution which is described by a 
tilted parabola. The normalized form of the parabola reads
\begin{equation*}
\widetilde J(\theta)=\frac{a+b\cos\theta+c\cos^{2}\theta}{2(a+c/3)}.
\end{equation*}
The linear
coefficient $b/2(a+c/3)$ can be projected out by defining a 
forward-backward asymmetry given by 
\begin{eqnarray*}
\mathcal{A}_{FB}(q^2)\\ =
\frac{ [\int_{0}^{1}-\int_{-1}^{0}] d\cos\theta\, d\Gamma/d\cos\theta}
     { [\int_{0}^{1}+\int_{-1}^{0}] d\cos\theta\, d\Gamma/d\cos\theta}\\ 
= \frac{b}{2(a+c/3)}
=\frac32 \frac{J_{6c}+2J_{6s}}{J_{\rm tot}}.
\end{eqnarray*}
The coefficient $c/2(a+c/3)$ of the quadratic contribution is obtained
by taking the second derivative of $\widetilde J(\theta)$. Accordingly, we
define a convexity parameter by writing 
\begin{equation*}
C_F^\tau(q^2) = \frac{d^{2}\widetilde J(\theta)}{d(\cos\theta)^{2}}
= \frac{c}{a+c/3} 
= \frac{6(J_{2c}+2J_{2s})}{J_{\rm tot}}.
\end{equation*}
\noindent
The $q^2$ dependence of $\mathcal{A}_{FB}$ is shown in Fig.~\ref{fig:AFB}. The coupling $V_L$ does not effect $\mathcal{A}_{FB}$ in both decays. In the case of the $\bar{B}^0\to D^{\ast}$ transition, the operators $\mathcal{O}_{V_R}$, $\mathcal{O}_{S_L}$, and $\mathcal{O}_{T_L}$ behave mostly similarly: they tend to decrease $\mathcal{A}_{FB}$ and shift the zero-crossing point to greater values than the SM one. However, the tensor operator can also increase $\mathcal{A}_{FB}$ in the high-$q^2$ region. In the case of the $\bar{B}^0\to D$ transition, the operator $\mathcal{O}_{V_R}$ does not affect $\mathcal{A}_{FB}$, the tensor operator $\mathcal{O}_{T_L}$ tends to lower $\mathcal{A}_{FB}$, and the scalar operator $\mathcal{O}_{S_L}$ thoroughly changes $\mathcal{A}_{FB}$: it can increase $\mathcal{A}_{FB}$ by up to $200\%$ and implies a zero-crossing point, which is impossible in the SM. This unique effect of $\mathcal{O}_{S_L}$ would clearly distinguish it from the other NP operators.

\begin{wrapfigure}{r}{0.58\textwidth}
\begin{tabular}{rr}
\includegraphics[scale=0.3]{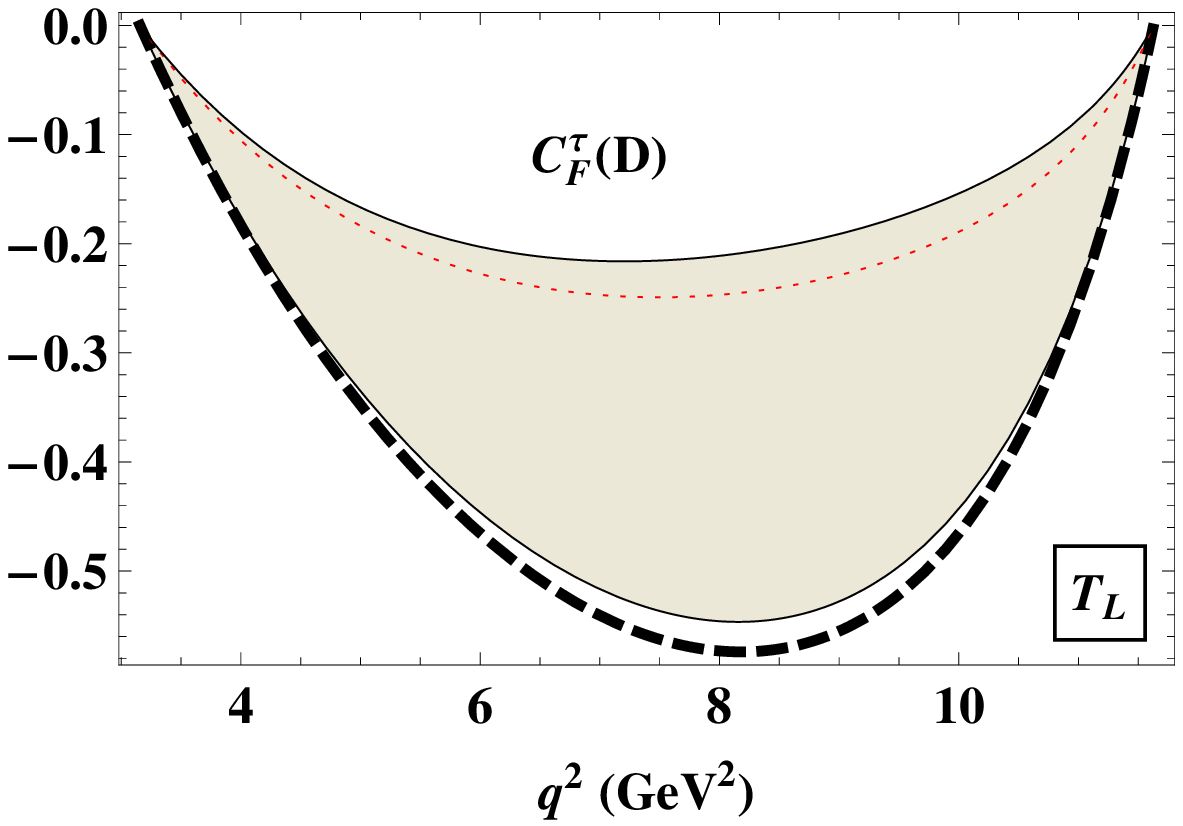}&
\includegraphics[scale=0.3]{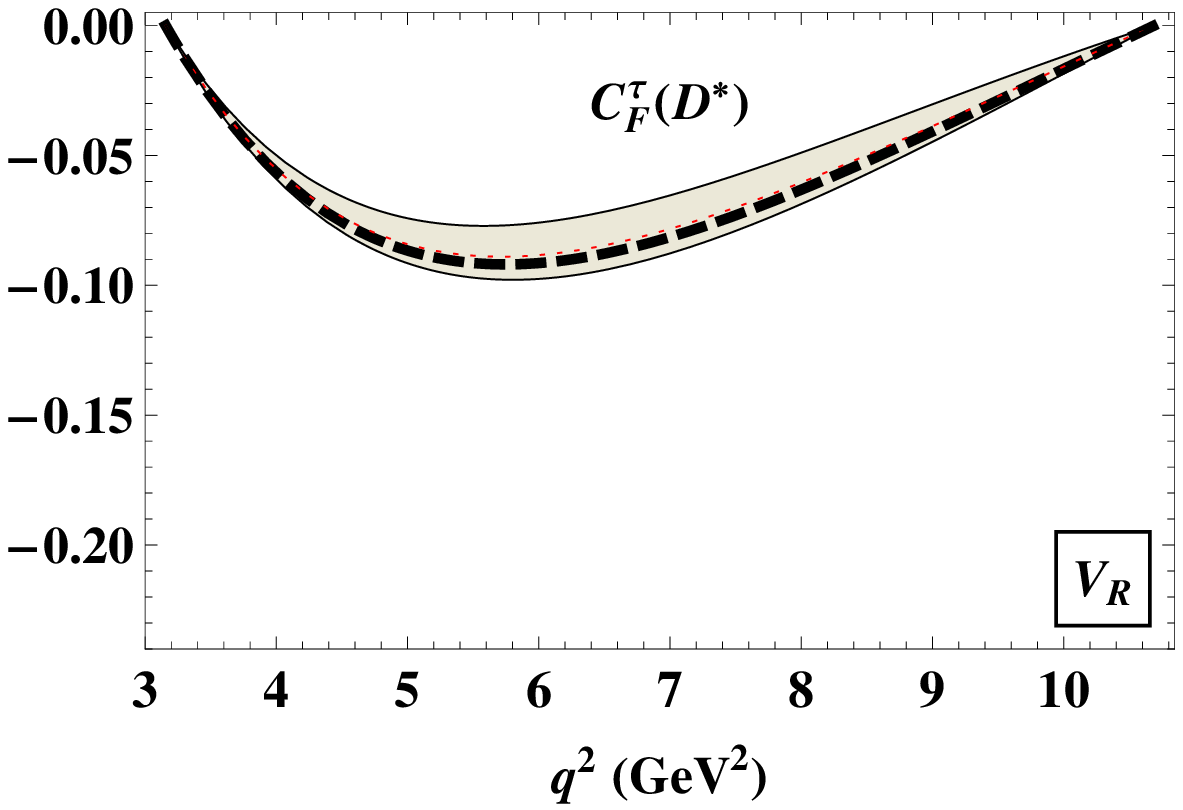}\\
\includegraphics[scale=0.3]{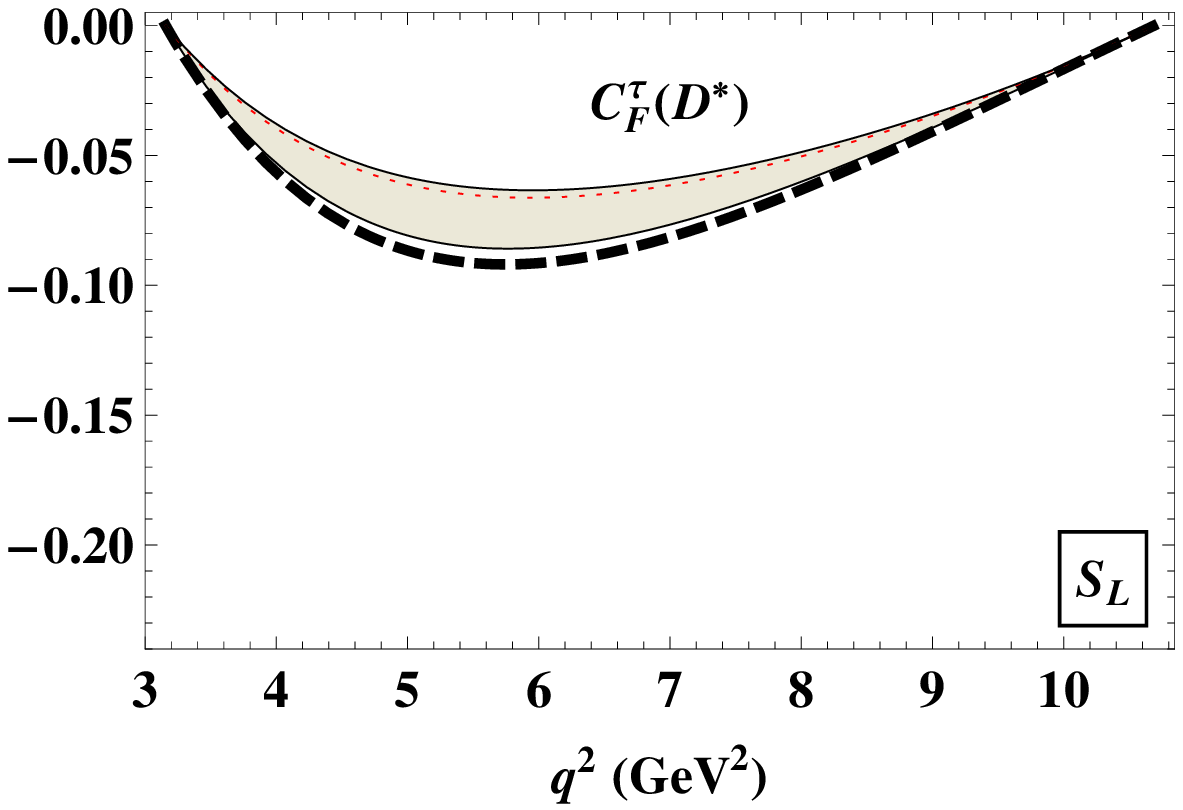}&
\includegraphics[scale=0.3]{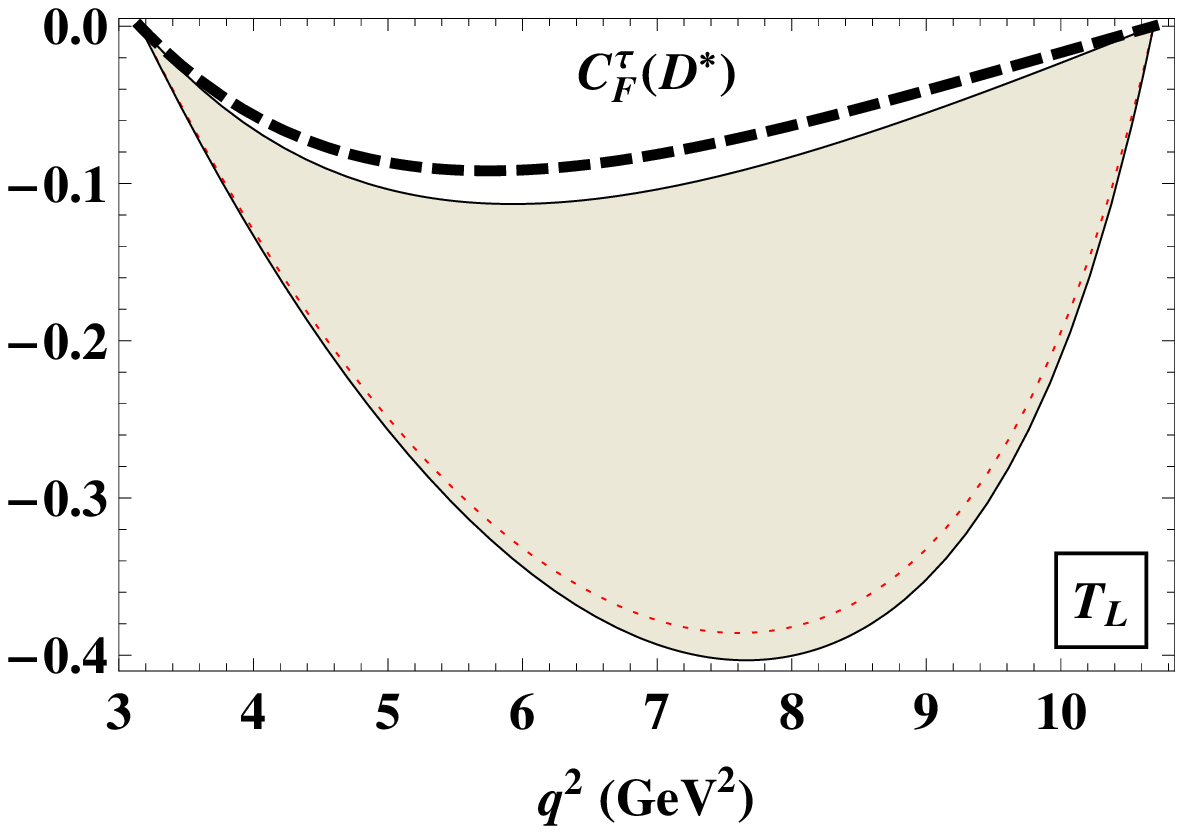}
\end{tabular}
\caption{Lepton-side convexity parameter $C_F^\tau(q^2)$.}
\label{fig:CFL}
\end{wrapfigure}

In Fig.~\ref{fig:CFL} we present 
the lepton-side
convexity parameter $C_F^\tau(q^2)$.
 While $C_F^\tau(D)$ is only sensitive to $\mathcal{O}_{T_L}$, $C_F^\tau(D^{\ast})$ is sensitive to $\mathcal{O}_{S_L}$, $\mathcal{O}_{V_R}$, and $\mathcal{O}_{T_L}$. 
Unlike $\mathcal{O}_{S_L}$, which can only increase $C_F^\tau(D^{\ast})$, the operator $\mathcal{O}_{T_L}$ can only lower the parameter. It is worth mentioning that $C_F^\tau(D)$ and $C_F^\tau(D^{\ast})$ are extremely sensitive to $\mathcal{O}_{T_L}$: it can change $C_F^\tau(D^{(\ast)})$ by a factor of 4 at $q^2\approx 7~\text{GeV}^2$.
%----------------------------------------------------------------------------
\subsection{The $\cos\theta^\ast$ distribution and the hadron-side convexity parameter}
By integrating the fourfold distribution over 
$\cos\theta$ and 
$\chi$ one obtains the hadron-side $\cos\theta^\ast$ distribution described 
by an untilted parabola (without a linear term). The normalized form of the  
$\cos\theta^\ast$ distribution
reads $\widetilde {J} (\theta^\ast)=(a'+c'\cos^{2}\theta^\ast)/2(a'+c'/3)$,
which can again be characterized by its convexity 
parameter given by
\begin{equation*}
C_F^h(q^2) = \frac{d^{2}\widetilde J(\theta^\ast)}{d(\cos\theta^{\ast})^{2}}
=\frac{c'}{a'+c'/3}=
\frac{3J_{1c}-J_{2c}-3J_{1s}+J_{2s}}{J_{\rm tot}/3}.
\end{equation*}
The $\cos\theta^\ast$ distribution can be written as
\begin{equation*}
\widetilde J(\theta^\ast)=3/4\cdot\left(2F_L(q^2)\cos^2\theta^\ast+F_T(q^2)\sin^2\theta^\ast\right),
\end{equation*}
where $F_L(q^2)$ and $F_T(q^2)$ are the polarization fractions of the $D^\ast$ meson and are defined as
\begin{equation*}
F_L(q^2)=J_L/(J_L+J_T),\qquad F_T(q^2)=J_T/(J_L+J_T),\qquad F_L(q^2)+F_T(q^2)=1.
\end{equation*}
\noindent
The hadron-side convexity parameter and the polarization fractions of the $D^\ast$ are related by
\begin{equation*} 
C_F^h(q^2)=3/2\cdot \left( 2F_L(q^2)-F_T(q^2) \right)=3/2\cdot \left(3F_L(q^2)-1\right).
\end{equation*}
\begin{figure}[htbp]
\begin{tabular}{ccc}
\includegraphics[scale=0.3]{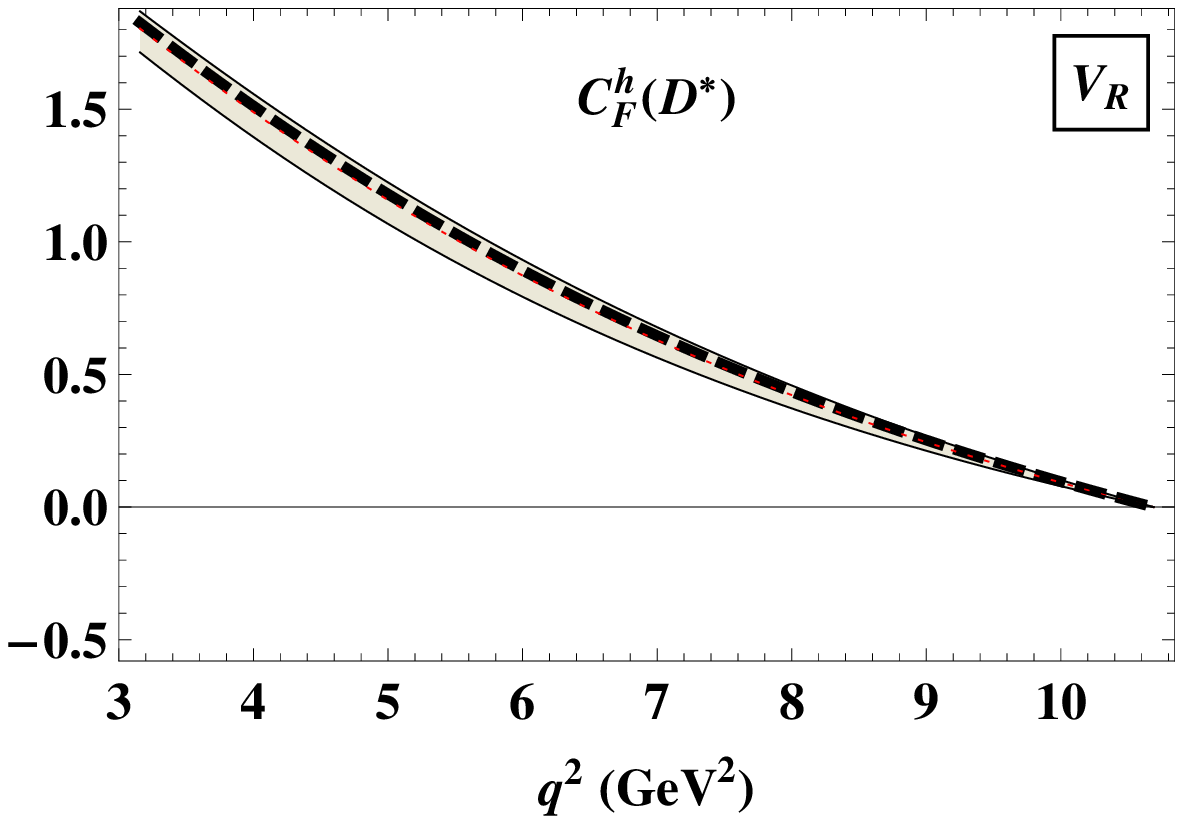}&
\includegraphics[scale=0.3]{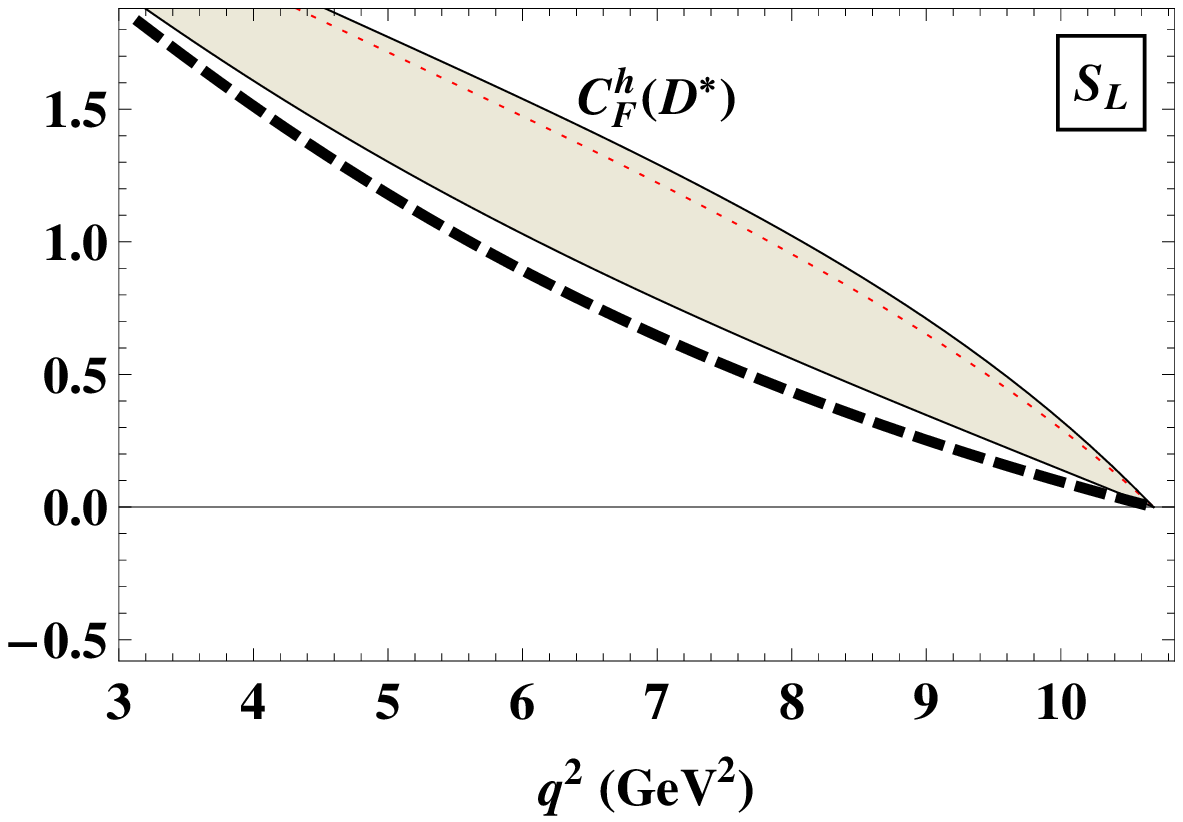}&
\includegraphics[scale=0.3]{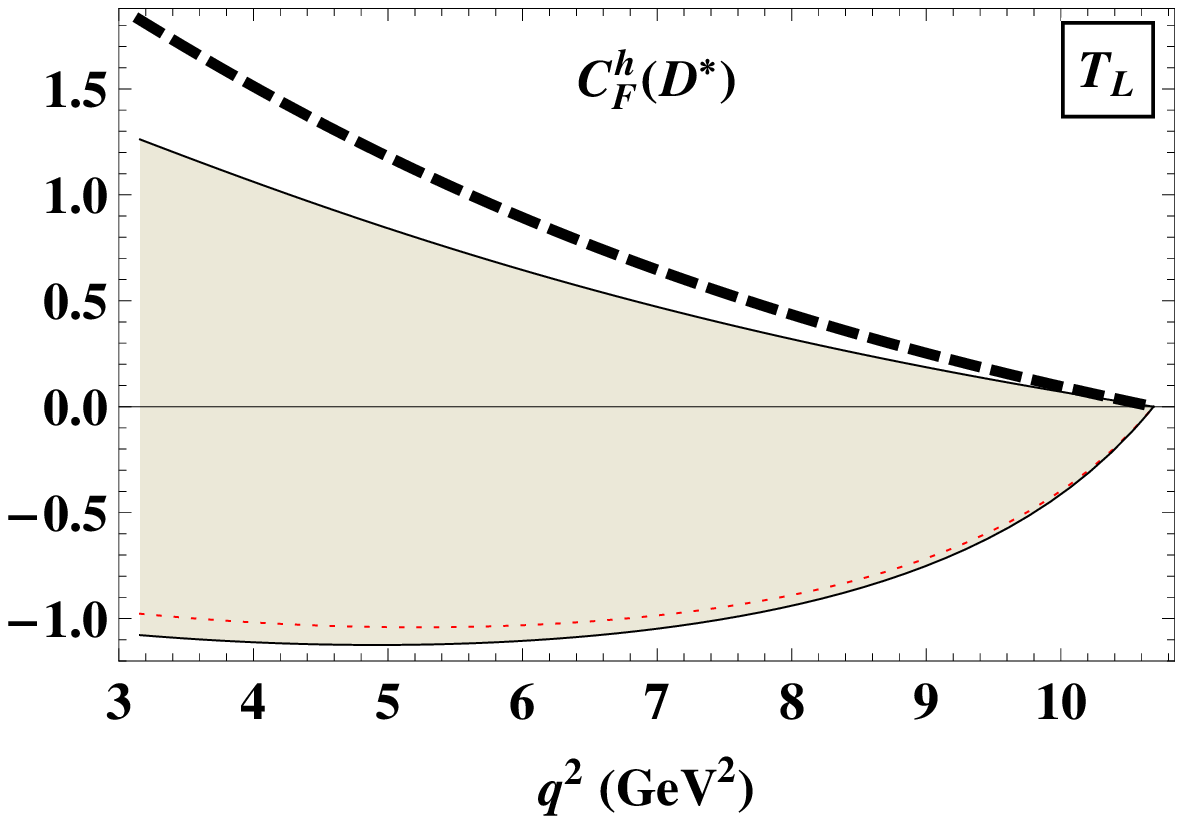}
\end{tabular}
\caption{Hadron-side convexity parameter $C_F^h(q^2)$. Notations are the same as in Fig.~\ref{fig:RD}.}
\label{fig:CFH}
\end{figure}
The effects of NP operators on the hadron-side convexity parameter $C_F^h(q^2)$ are shown in Fig.~\ref{fig:CFH}. Each NP operator can change $C_F^h(q^2)$ in a unique way: the vector operator $\mathcal{O}_{V_R}$ almost does nothing to the parameter; the scalar operator $\mathcal{O}_{S_L}$ increases the parameter by about $50\%$ nearly in the whole range of $q^2$; the tensor operator $\mathcal{O}_{T_L}$ lowers the parameter (by up to $200\%$ at low $q^2$), and it also allows negative values of $C_F^h(q^2)$, which are impossible in the SM.
%----------------------------------------------------------------------------
\subsection{The $\chi$ distribution and the trigonometric moments}
By integrating the fourfold distribution over $\cos\theta$ and $\cos\theta^\ast$,
one obtains the $\chi$ distribution whose normalized form reads
\begin{equation*}
\widetilde J^{(I)}(\chi)=\Big[1+A_C^{(1)}(q^2)\cos 2\chi+A_T^{(1)}(q^2)\sin 2\chi\Big]/(2\pi),
\end{equation*}
where $A_C^{(1)}(q^2)=4J_3/J_{\rm tot}$ and $A_T^{(1)}(q^2)=4J_9/J_{\rm tot}$. Besides, one can also define other angular distributions in the angular variable $\chi$ as follows:
\begin{eqnarray*}
J^{(II)}(\chi)&=&\Big[\int_0^1-\int_{-1}^0\Big]d\cos \theta^\ast\int_{-1}^1d\cos \theta\frac{d^4\Gamma}
     {dq^2 d\cos\theta d\chi d\cos\theta^\ast}, 
\\
J^{(III)}(\chi)&=&\Big[\int_0^1-\int_{-1}^0\Big]d\cos \theta^\ast\Big[\int_0^1-\int_{-1}^0\Big]d\cos \theta\frac{d^4\Gamma}
     {dq^2 d\cos\theta d\chi d\cos\theta^\ast}.     
\end{eqnarray*}
The normalized forms of these distributions read
\begin{eqnarray*}
\widetilde J^{(II)}(\chi)&=&\frac14\Big[A_C^{(2)}(q^2)\cos\chi+A_T^{(2)}(q^2)\sin\chi\Big],\\
\widetilde J^{(III)}(\chi)&=&\frac{2}{3\pi}\Big[A_C^{(3)}(q^2)\cos\chi+A_T^{(3)}(q^2)\sin\chi\Big],
\end{eqnarray*}
where 
$A_C^{(2)}(q^2)=3J_5/J_{\rm tot},\quad A_T^{(2)}(q^2)=3J_7/J_{\rm tot},\quad A_C^{(3)}(q^2)=3J_4/J_{\rm tot},\quad A_T^{(3)}(q^2)=3J_8/J_{\rm tot}$.

Another method to project the coefficient functions $J_i$ $(i=3,4,5,7,8,9)$ out from the fourfold angular decay distribution is to take the
appropriate trigonometric moments of the normalized decay distribution
$\widetilde J(\theta^\ast,\theta,\chi)$~\cite{Ivanov:2015tru}. The trigonometric moments are defined by
\begin{equation*}
W_{i} = \int d\cos\theta d\cos\theta^\ast d\chi
M_{i}(\theta^\ast,\theta,\chi)\widetilde J(\theta^\ast,\theta,\chi) 
\equiv  \left\langle M_{i}(\theta^\ast,\theta,\chi) \right\rangle,
\end{equation*}
where $M_{i}(\theta^\ast,\theta,\chi)$ defines the trigonometric moment that 
is being taken.
One finds 
\begin{eqnarray*}
W_T(q^2) &\equiv& \left\langle \cos 2\chi \right\rangle  
= 2\cdot (J_3/J_{\rm tot}) =(1/2) \cdot A_C^{(1)}(q^2),\\
W_{IT}(q^2) &\equiv& \left\langle \sin 2\chi \right\rangle  
= 2\cdot(J_9/J_{\rm tot}) = (1/2)\cdot A_T^{(1)}(q^2),\\
W_A(q^2) &\equiv& \left\langle \sin\theta\cos\theta^{\ast}\cos \chi \right\rangle 
= (3\pi /8)\cdot (J_5/J_{\rm tot}) = (\pi /8)\cdot A_C^{(2)}(q^2),\\
W_{IA}(q^2) &\equiv& \left\langle \sin\theta\cos\theta^{\ast}\sin \chi \right\rangle 
= (3\pi/8)\cdot (J_7/J_{\rm tot}) = (\pi/8) \cdot A_T^{(2)}(q^2),\\
W_I(q^2) &\equiv& \left\langle   
\cos\theta\cos\theta^{\ast}\cos \chi \right\rangle  
= (9\pi^2/128)\cdot(J_4/J_{\rm tot}) = (3\pi^2/128)\cdot A_C^{(3)}(q^2),\\
W_{II}(q^2) &\equiv& \left\langle   
\cos\theta\cos\theta^{\ast}\sin \chi \right\rangle  
=(9\pi^2/128)\cdot(J_8/J_{\rm tot}) = (3\pi^2/128)\cdot A_T^{(3)}(q^2).
\end{eqnarray*}
\begin{figure}[htbp]
\begin{tabular}{ccc}
\includegraphics[scale=0.28]{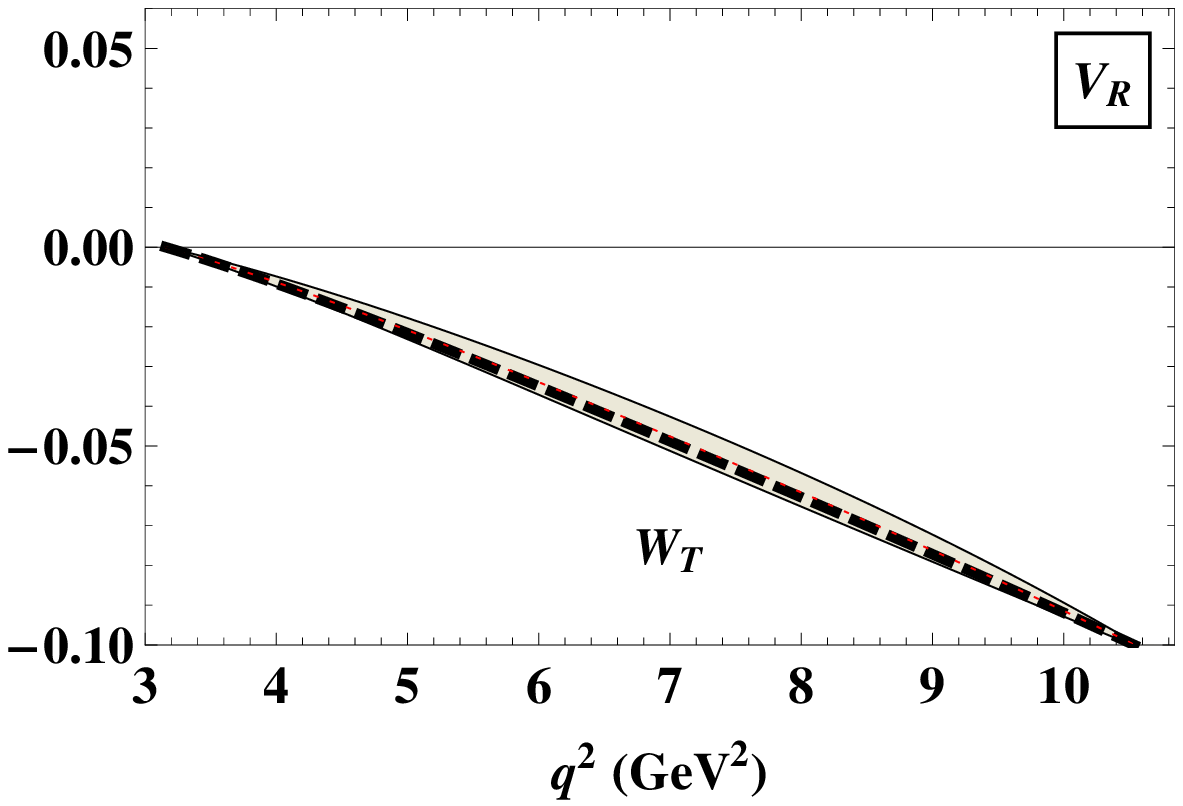}&
\includegraphics[scale=0.28]{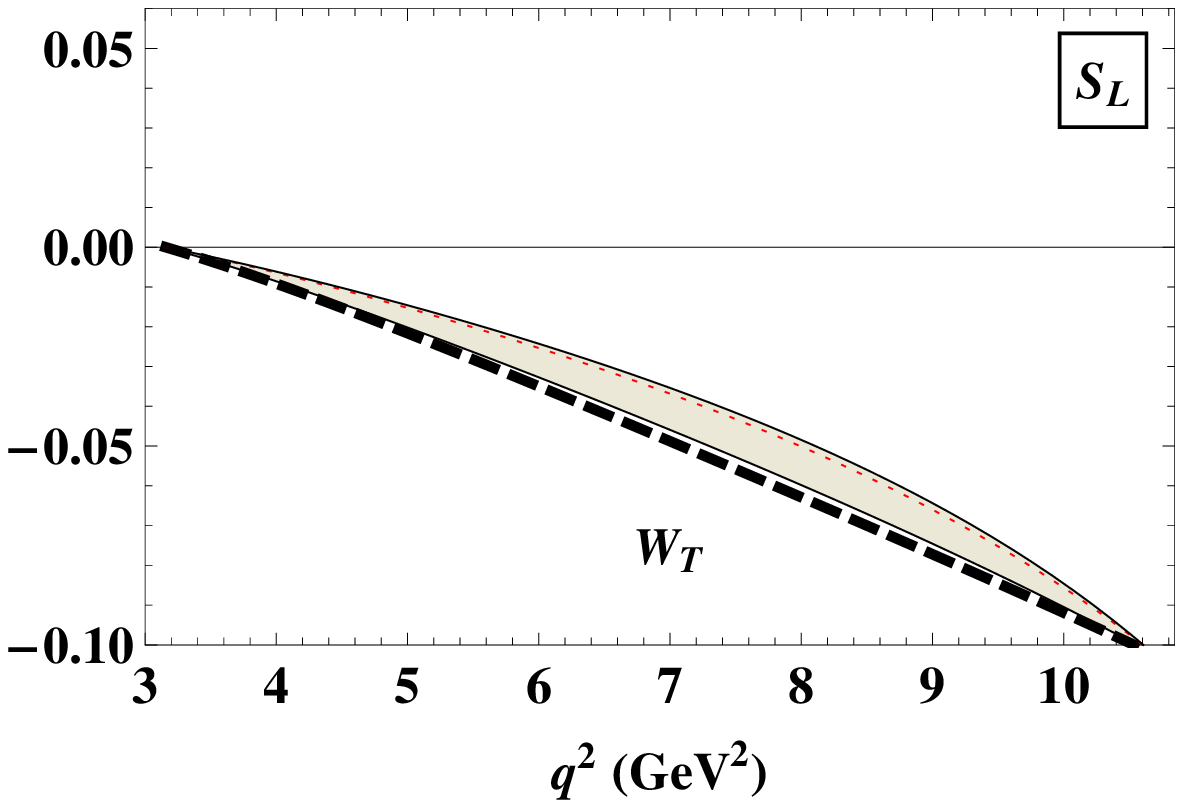}&
\includegraphics[scale=0.28]{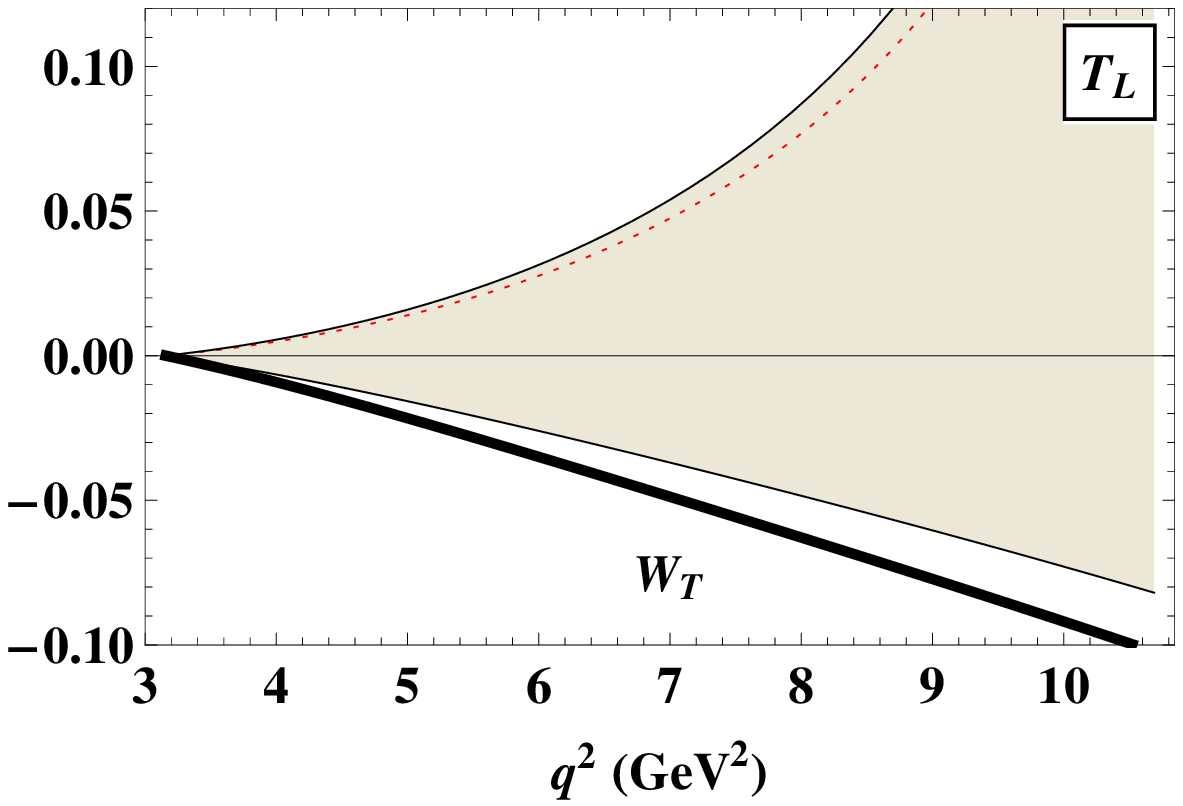}\\
\includegraphics[scale=0.28]{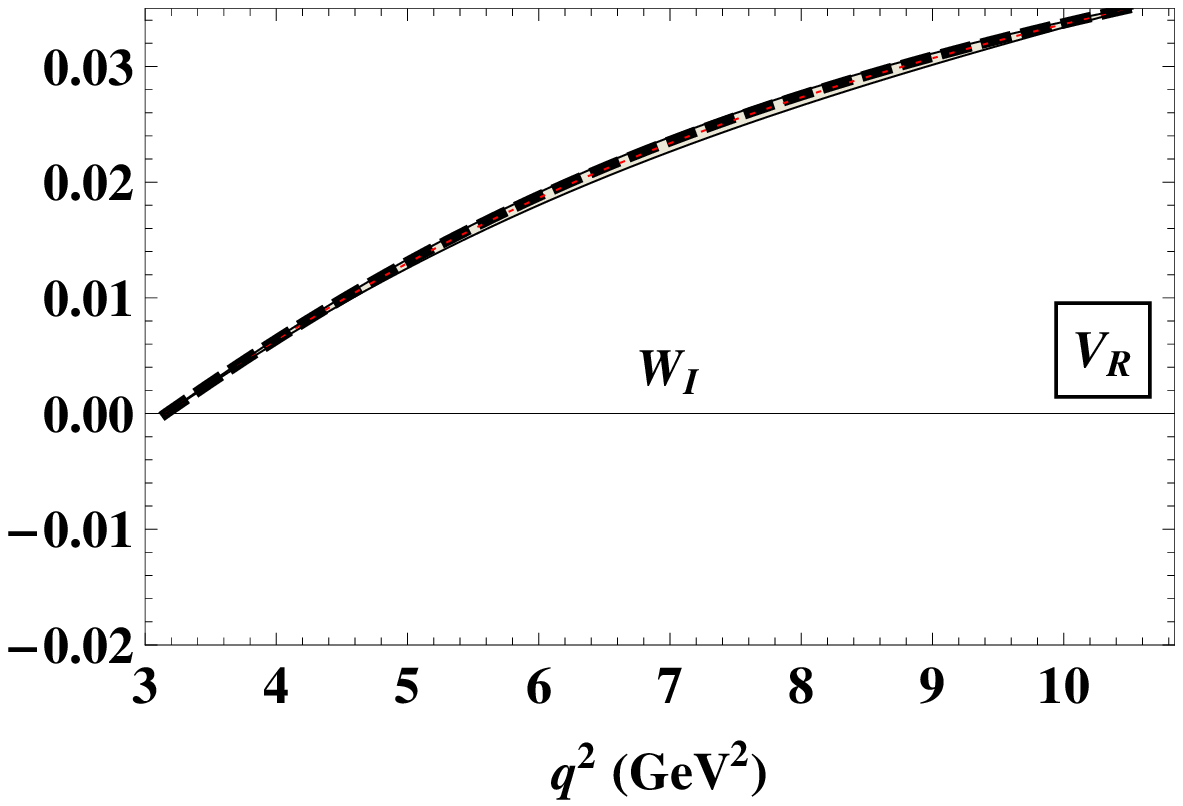}&
\includegraphics[scale=0.28]{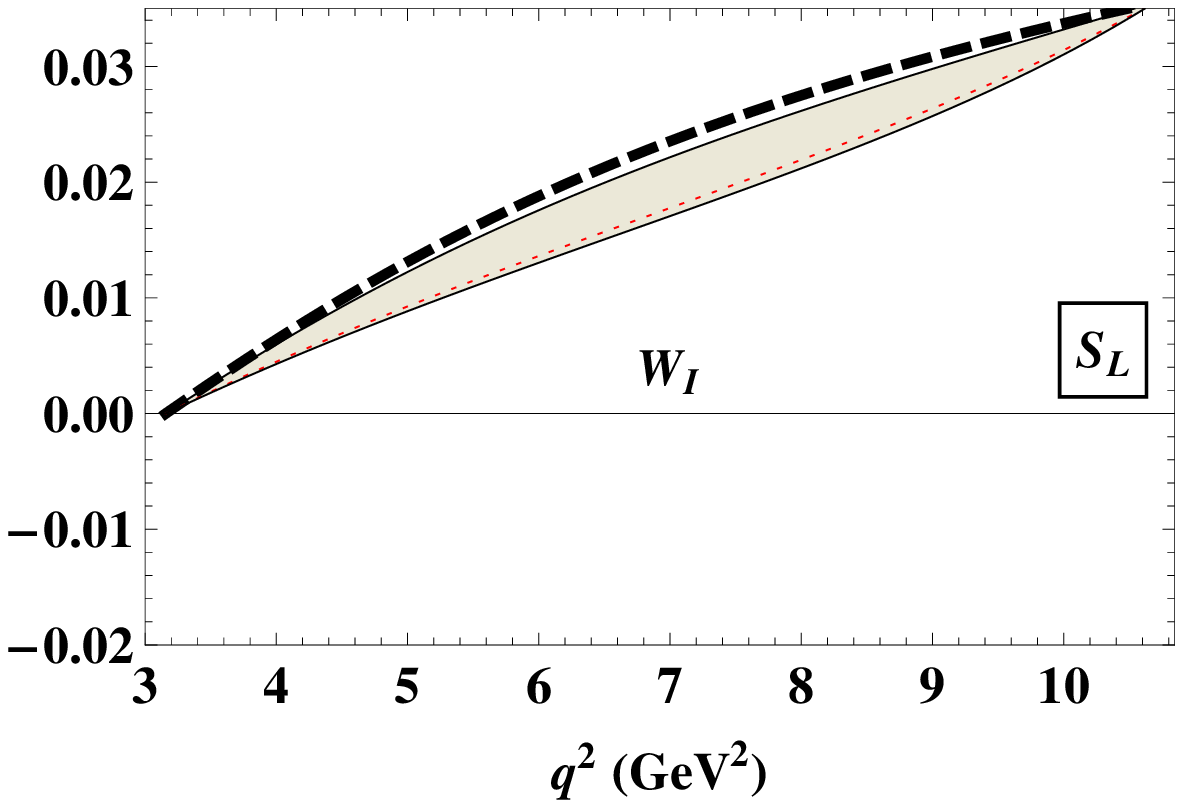}&
\includegraphics[scale=0.28]{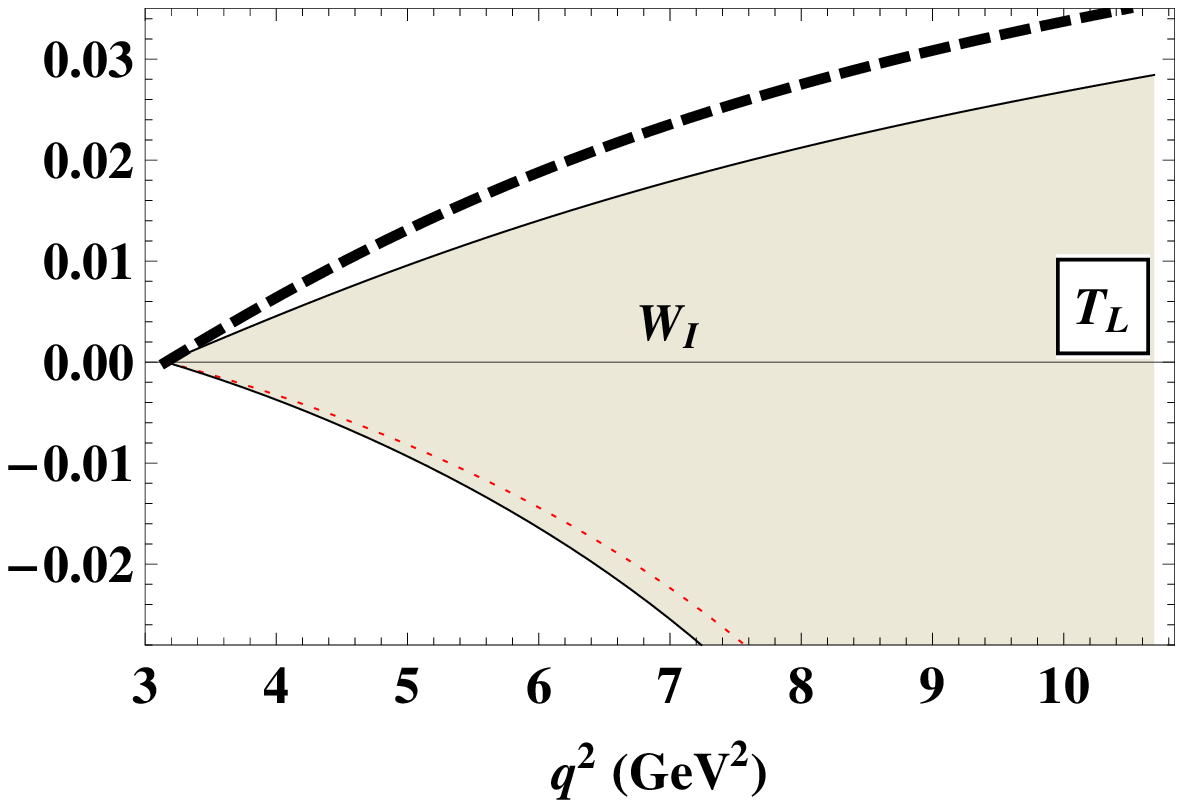}\\
\includegraphics[scale=0.28]{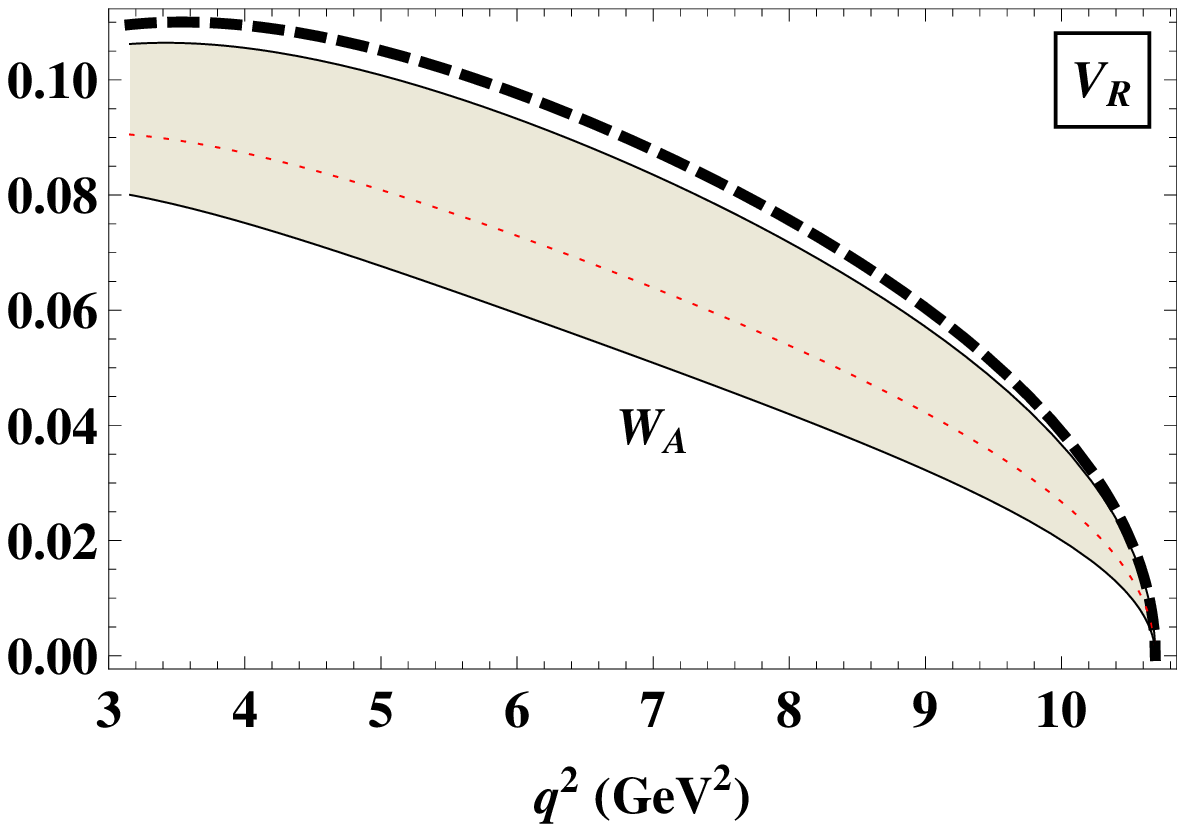}&
\includegraphics[scale=0.28]{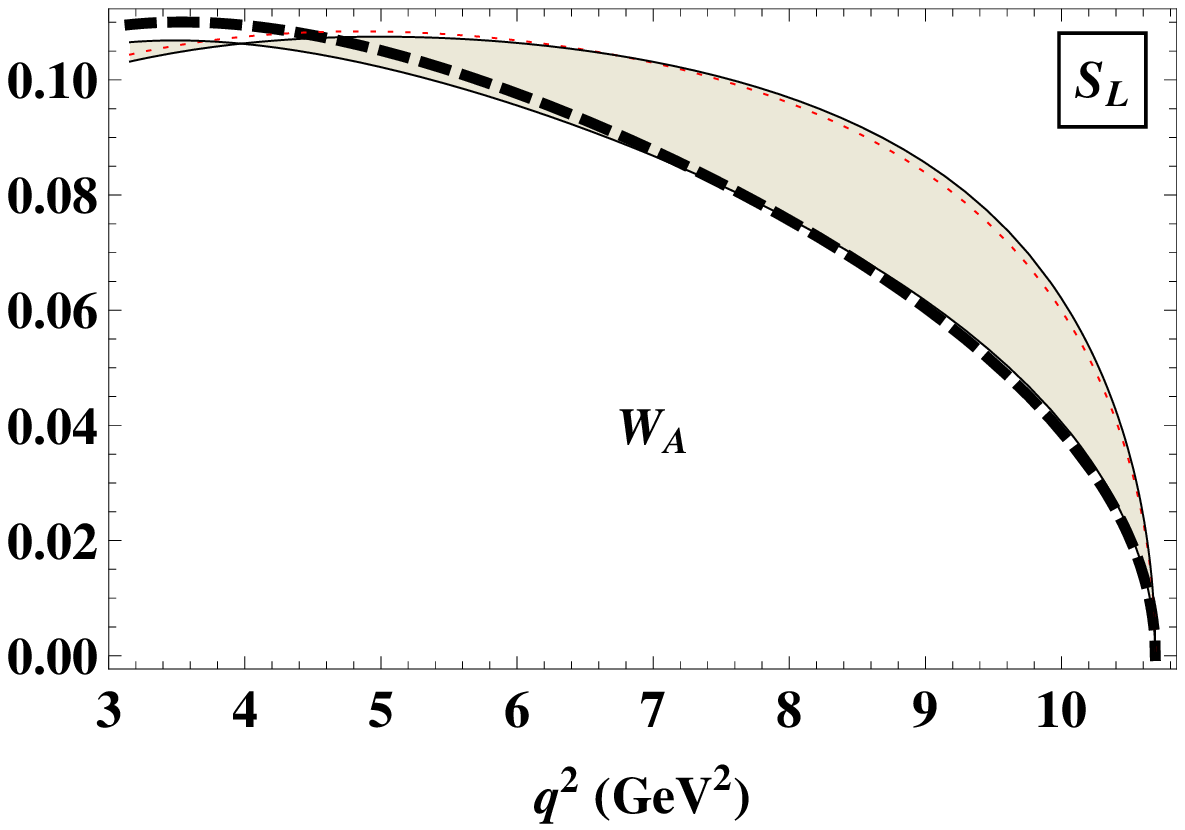}&
\includegraphics[scale=0.28]{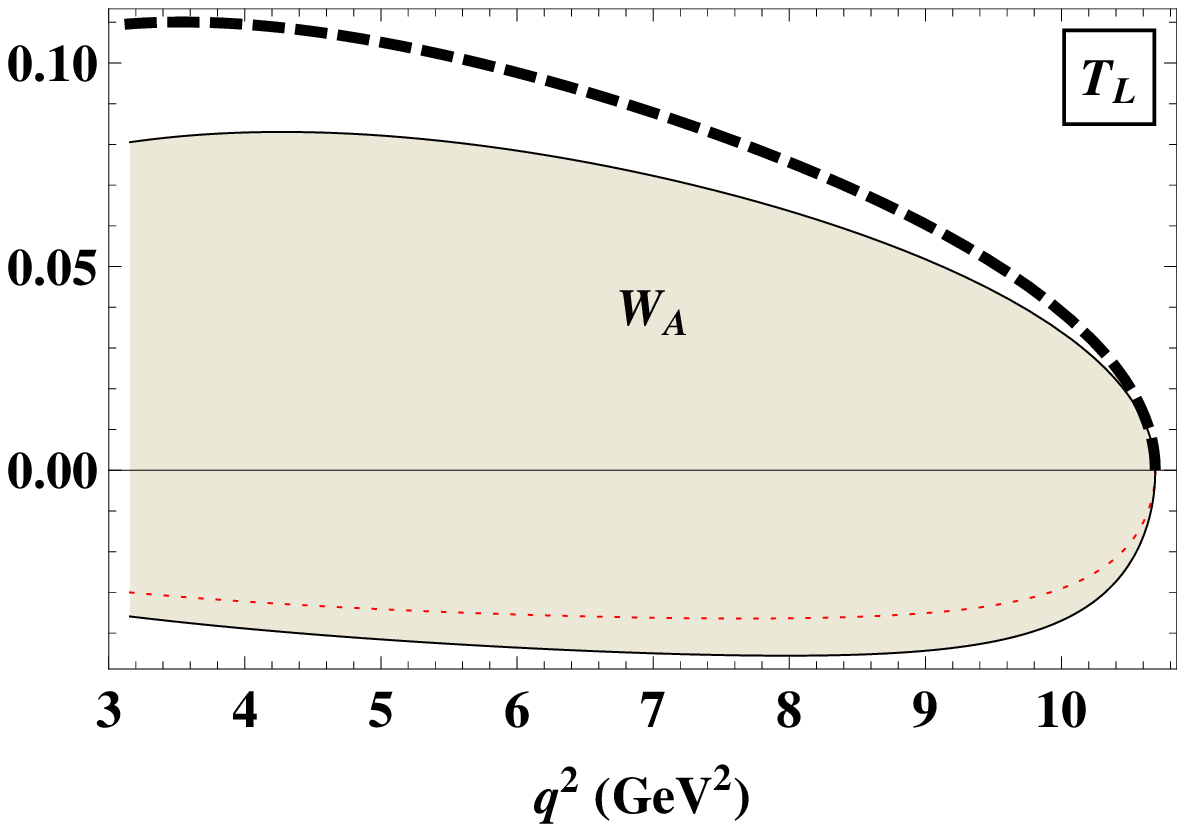}\\
\includegraphics[scale=0.28]{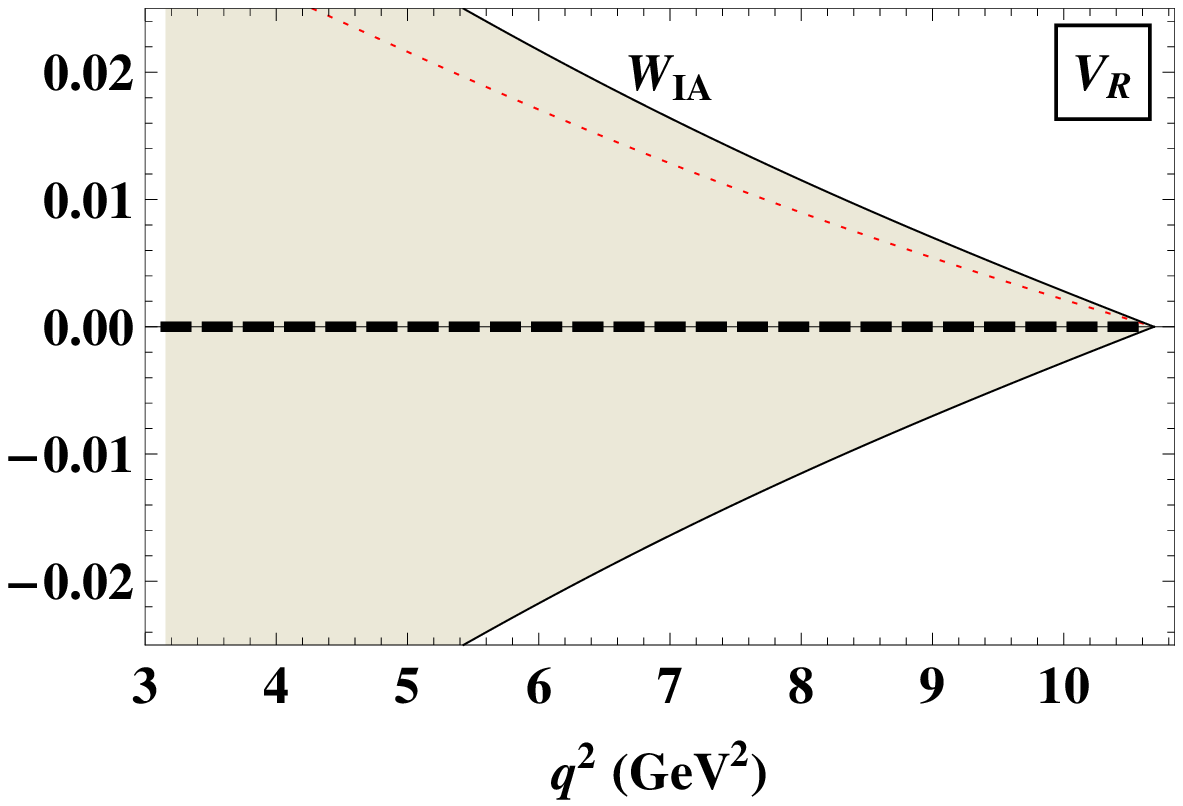}&
\includegraphics[scale=0.28]{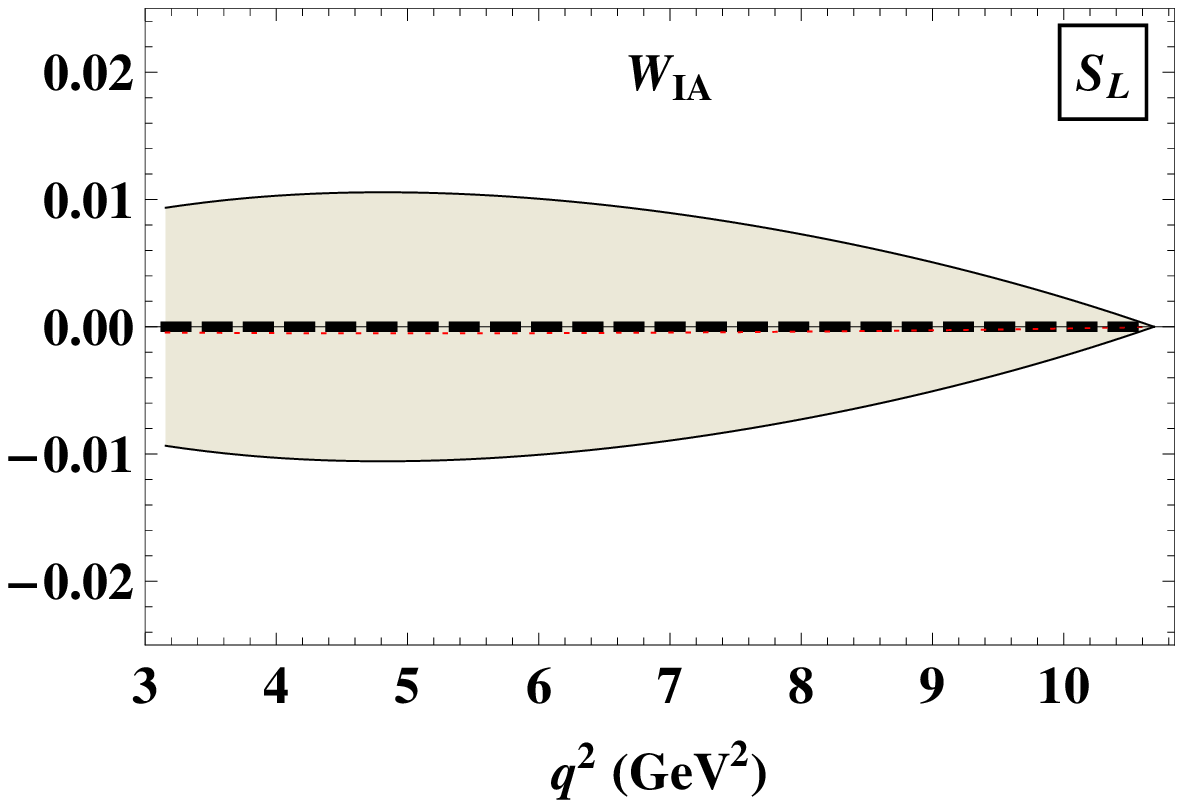}&
\includegraphics[scale=0.28]{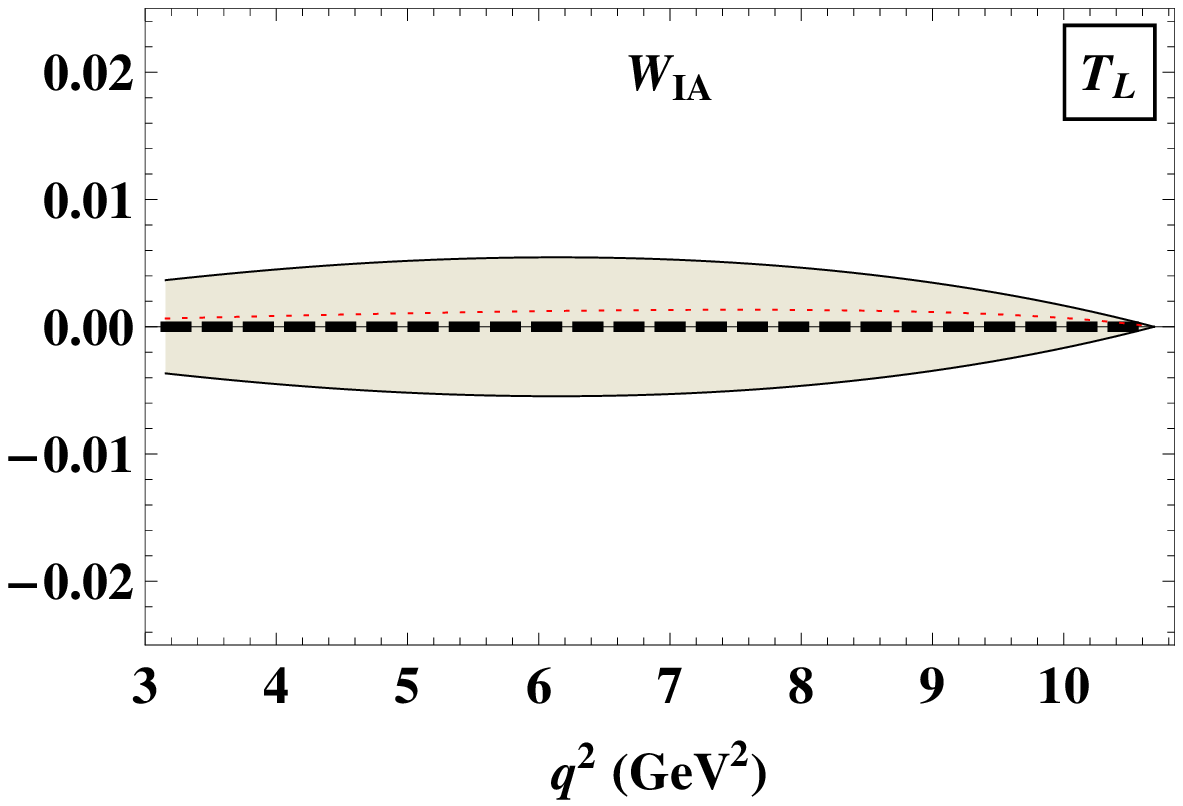}
\end{tabular}
\caption{Trigonometric moments $W_T(q^2)$, $W_I(q^2)$, $W_A(q^2)$, and $W_{IA}(q^2)$. Notations are the same as in Fig.~\ref{fig:RD}.}
\label{fig:WTAI}
\end{figure}
In Fig.~\ref{fig:WTAI} we show the $q^2$ dependence of the trigonometric moments $W_T(q^2)$, $W_I(q^2)$, $W_A(q^2)$, and $W_{IA}(q^2)$. The moments $W_T(q^2)$ and  $W_I(q^2)$ are almost insensitive to $\mathcal{O}_{V_R}$ but highly sensitive to $\mathcal{O}_{T_L}$. The scalar and tensor operators  are likely to raise $W_T(q^2)$ and to lower $W_I(q^2)$ in general. The moment $W_A(q^2)$ shows great sensitivity to $\mathcal{O}_{V_R}$, $\mathcal{O}_{S_L}$, and $\mathcal{O}_{T_L}$. Both $\mathcal{O}_{V_R}$ and $\mathcal{O}_{T_L}$ tend to decrease $W_A(q^2)$ while $\mathcal{O}_{S_L}$ tries to do the opposite. It is worth noting that all three moments $W_T(q^2)$, $W_I(q^2)$, and $W_A(q^2)$ are extremely sensitive to $\mathcal{O}_{T_L}$ and their sign can change in the presence of $\mathcal{O}_{T_L}$. Regarding the moment $W_{IA}(q^2)$, the three operators act in the same manner: they can change $W_{IA}(q^2)$ in both directions, and the sensitivity is maximal in the case of $\mathcal{O}_{V_R}$.
\begin{figure}[htbp]
\begin{tabular}{ccc}
\includegraphics[scale=0.35]{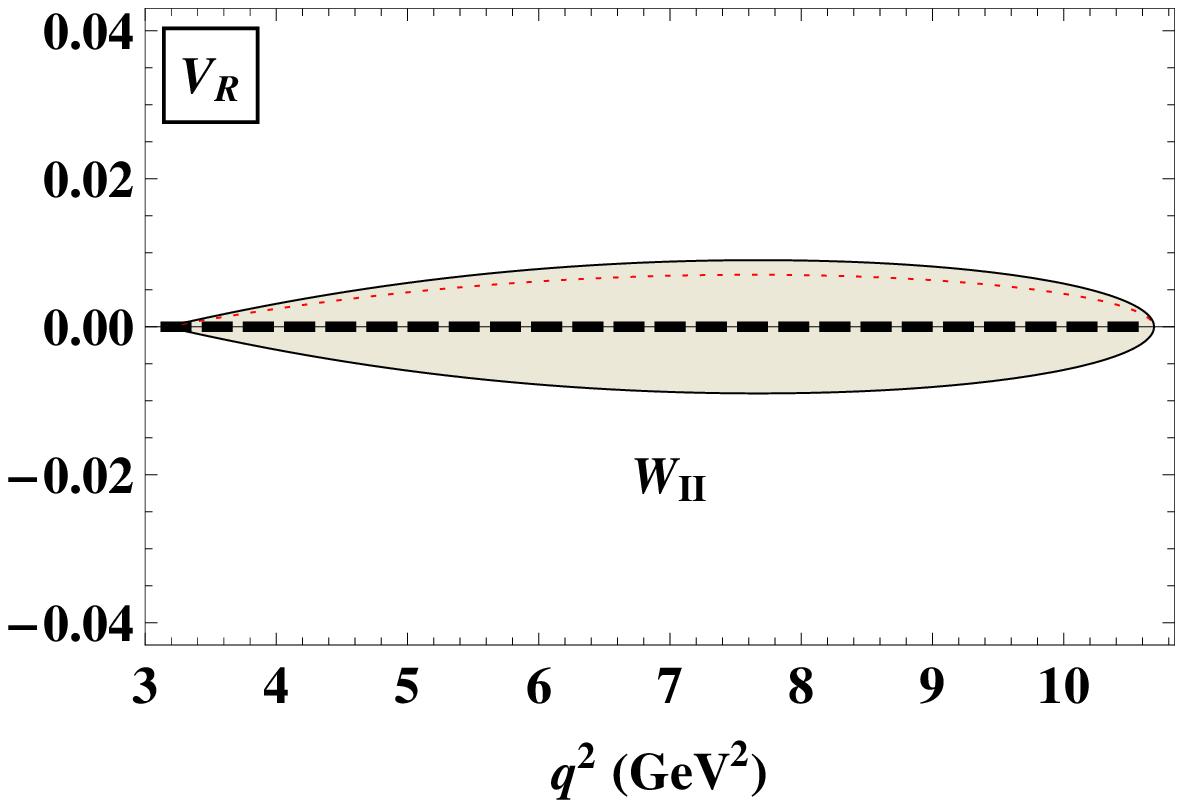}&
\includegraphics[scale=0.35]{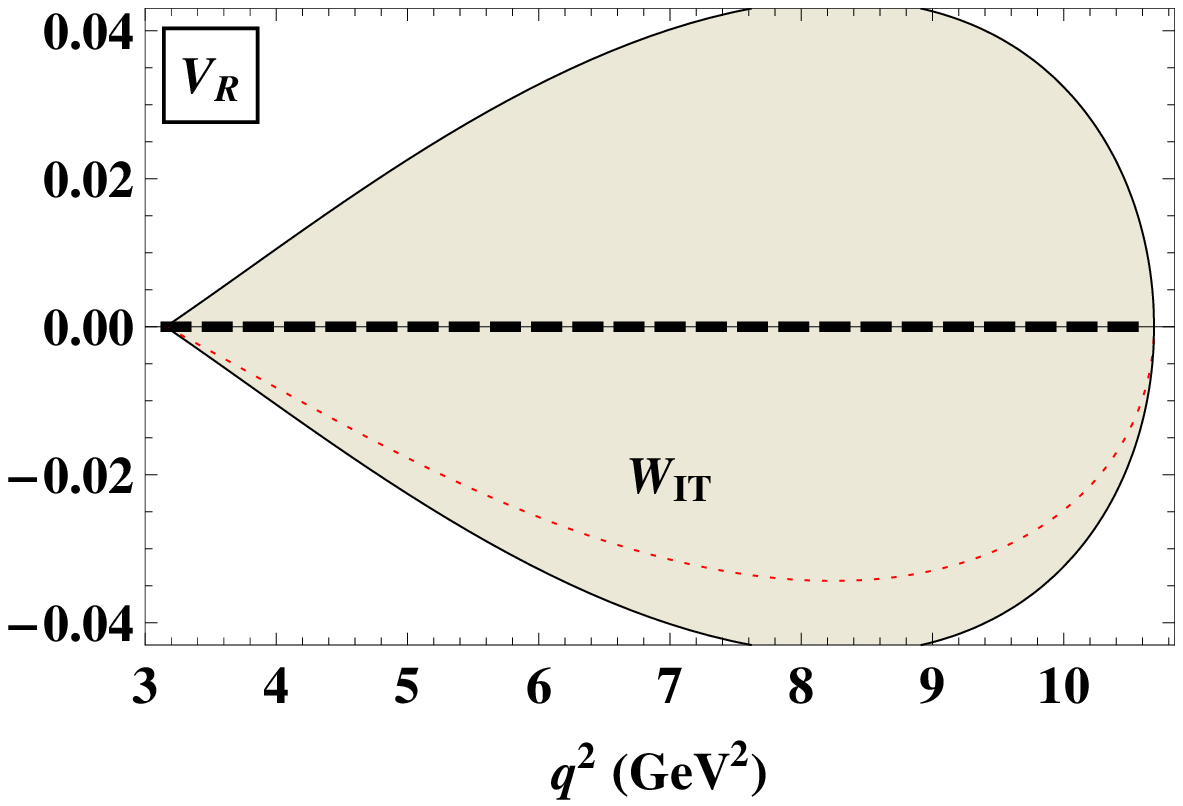}&
\includegraphics[scale=0.35]{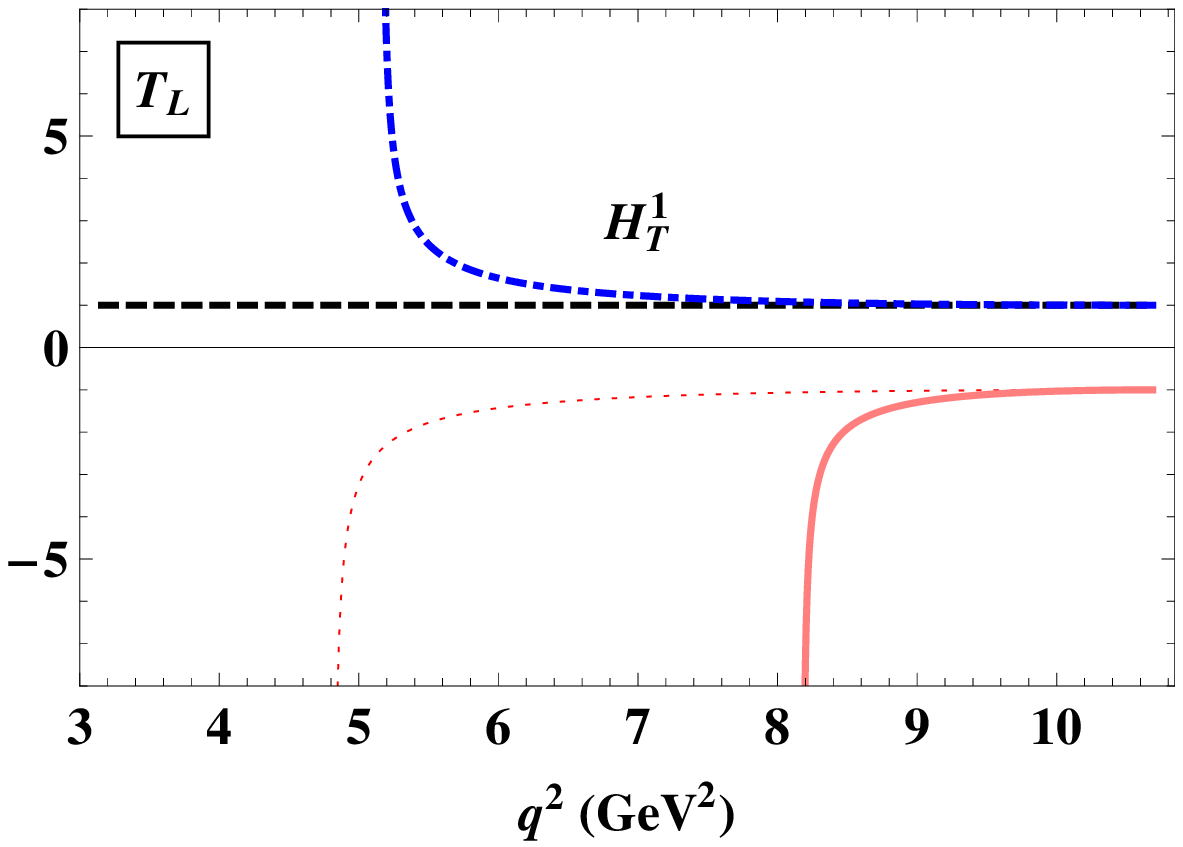}
\end{tabular}
\caption{Trigonometric moments $W_{II}(q^2)$ (left) and $W_{IT}(q^2)$ (center), and  the optimal angular observable $H_T^{(1)}(q^2)$ (right). The black dashed lines are the SM prediction. The red dotted lines represent the best-fit values. The blue dot-dashed line and the pink solid line in the right plot are the predictions for $T_L=0.04-0.17i$ and $T_L=0.18+0.23i$, respectively.}
\label{fig:WIIT}
\end{figure}
The trigonometric moments $W_{II}(q^2)$ and $W_{IT}(q^2)$ are equal to zero in the SM and obtain a nonzero contribution only from the right-chiral vector operator
$\mathcal{O}_{V_R}$, as depicted in Fig.~\ref{fig:WIIT}. Both moments are proportional to the imaginary part of $V_R$ and the effect of $\mathcal{O}_{V_R}$ cancels in their ratio.
%----------------------------------------------------------------

One can also consider certain combinations of angular observables where the form factor dependence drops out (at least in most NP scenarios)~\cite{Feldmann:2015xsa}. As a demonstration, we consider the optimized observable 
\begin{equation*}
H_T^{(1)} = \sqrt{2}J_4/\sqrt{-J_{2c}(2J_{2s}-J_3)},
\end{equation*}
which is equal to one not only in the SM but also in all NP scenarios except the tensor one, as shown in the right plot of Fig.~\ref{fig:WIIT}. Therefore $H_T^{(1)}(q^2)$ plays a prominent role in confirming the appearance of the tensor operator $\mathcal{O}_{T_L}$ in the decay $\bar{B}^0 \to D^{\ast} \tau^- \bar{\nu}_{\tau}$.
%----------------------------------------------------------------------------
\section{Tau polarization as probe for NP}
Recently, the Belle collaboration has reported
on the first measurement of the longitudinal polarization of the tau lepton in the decay
$\bar{B}^0 \to D^\ast\tau^-\bar\nu_{\tau}$ with the result $P_L^\tau = -0.38 \pm 0.51(\text{stat.})
^{+0.21}_{-0.16} (\text{syst.})$~\cite{Belle}. The errors are quite
large but this pioneering measurement has opened a new window on the analysis of the dynamics of the
semileptonic $B\to D^{(*)}$ transitions. The hope is that, with
the Belle II super-B factory nearing completion, more precise values of the polarization can be achieved in the future, which would shed more light on the
search for possible NP in these decays.

In a recent paper~\cite{Ivanov:2017mrj} we have studied the longitudinal ($P_L$), transverse ($P_T$), and normal ($P_N$) polarization components of the $\tau^-$ in $\bar{B}^0 \to D^{(\ast)}\tau^-\bar\nu_{\tau}$ and clarified their roles in the search for NP. In~\cite{Ivanov:2017mrj} one can find
the $q^2$ dependence of the $\tau^-$ polarizations in the presence of NP operators, which bears powerful information for discriminating between different NP scenarios. For example, it can be used to perform a bin-by-bin analysis to probe NP in different $q^2$ regions. One can also calculate the average polarizations over the whole $q^2$ region. 
The predictions for the mean polarizations are summarized in Table~\ref{tab:pol-average}. 
%%%%%%%%%%%%%%%
\begin{table}[htbp] 
\begin{center}
\begin{tabular}{|l|cccc|c|}
\hline
\multicolumn{5}{|c|}{ $\bar{B}^0\to D$ }\\
\hline
&\quad  $<P_{L}^D>$\qquad 
&\quad  $<P_{T}^D>$ \qquad   
&\quad  $<P_{N}^D>$ \qquad  
&\quad  $<|\vec P^D|>$ \qquad
\\
\hline
SM (CCQM) &\quad $0.33$\quad &\quad $0.84$ \quad &\quad $0$\quad&\quad $0.91$\quad\\
$S_L$ 
&\quad $(0.36,0.67)$\quad
&\quad $(-0.68,0.33)$\quad
&\quad $(-0.76,0.76)$\quad
&\quad $(0.89,0.96)$\quad
\\ 
$T_L$
&\quad $(0.13,0.31)$\quad
&\quad $(0.78,0.83)$\quad
&\quad $(-0.17,0.17)$\quad
&\quad $(0.79,0.90)$\quad\\
\hline
\multicolumn{5}{|c|}{ $\bar{B}^0\to D^\ast$ }\\
\hline
&\quad  $<P_{L}^{D^\ast}>$ \qquad 
&\quad  $<P_{T}^{D^\ast}>$ \qquad   
&\quad  $<P_{N}^{D^\ast}>$ \qquad  
&\quad $<|\vec P^{D^\ast}|>$\qquad
\\
\hline
SM (CCQM)&\quad -0.50\quad &\quad 0.46\quad &\quad 0 \quad &\quad 0.71\quad\\
$S_L$
&\quad $(-0.40,-0.14)$\quad
&\quad $(0.47,0.62)$\quad
&\quad $(-0.20,0.20)$\quad
&\quad $(0.69,0.70)$\quad
\\
$T_L$
&\quad $(-0.36,0.24)$\quad
&\quad $(-0.61,0.26)$\quad
&\quad $(-0.17,0.17)$\quad
&\quad $(0.23,0.69)$\quad
\\
$V_R$
&\quad $-0.50$\quad
&\quad $(0.32,0.43)$\quad
&\quad $0$\quad
&\quad $(0.48,0.67)$\quad
\\
\hline
\end{tabular}
\caption{$q^{2}$ averages of the polarization components and the total polarization. The two rows labeled by SM (CCQM) contain our predictions within the SM with form factors calculated in the CCQM. The predicted intervals for the observables in the presence of NP are given in correspondence with the $2\sigma$ allowed regions of the NP couplings depicted in Fig.~\ref{fig:constraint}.}
\label{tab:pol-average}
\end{center}
\end{table}
%%%%%%
One sees that the $\tau^-$ polarization components in $\bar{B}^0 \to D\tau^-\bar\nu_{\tau}$ are extremely sensitive to $S_L$. When $S_L$ is present, $\langle P_L^D\rangle$ can be as large as $0.67$, $\langle P_T^D\rangle$ can reach $-0.68$, and $\langle P_N^D\rangle$ can even reach $\pm 0.76$. It is interesting to note that if one measures $\langle P_L^D\rangle$ and finds any excess over the SM value, it would be a clear sign of $S_L$. Meanwhile, the $\tau^-$ longitudinal and transverse polarization components in $\bar{B}^0 \to D^\ast\tau^-\bar\nu_{\tau}$ are more sensitive to $T_L$. The coupling $T_L$ can enhance $\langle P_L^{D^\ast}\rangle$ from the SM value of $-0.50$ up to $0.24$, or lower $\langle P_T^{D^\ast}\rangle$ from $0.46$ down to $-0.61$. Notably, the average transverse polarization $\langle P_T^D\rangle$ is almost insensitive to $T_L$ in comparison with $S_L$. When $T_L$ is present, one finds $0.78\leq\langle P_T^D\rangle\leq 0.83$, which is almost the same as the SM value $\langle P_T^D\rangle=0.84$. In contrast, if $S_L$ is present, one has $-0.68\leq\langle P_T^D\rangle\leq 0.33$, which is much lower than the SM prediction. This unique property of $\langle P_T^D\rangle$ may play a very important role in probing the scalar coupling $S_L$. It is also interesting to note that the average total polarization $<|\vec P^{D^\ast}|>$ is almost insensitive to $S_L$. 
%%%%%%%%%%%%%%%%%%
%%%%%%%%%%%%%%%%%%
\section{Summary and discussion}
We have provided a thorough analysis of possible NP in the decays $\bar{B}^0 \to D^{(\ast)}\tau^-\bar\nu_{\tau}$ using the form factors obtained from our covariant quark model. Starting with a general effective Hamiltonian including NP operators, we have derived the full angular distribution and defined a large set of physical observables. Assuming NP only affects leptons of the third generation and only one NP operator appears at a time, we have gained the allowed regions of NP couplings based on recent measurements at $B$ factories, and studied their effects on the observables. It has turned out that the current experimental data of $R(D)$ and $R(D^\ast)$ prefer the operators $\mathcal{O}_{S_L}$ and $\mathcal{O}_{V_{L,R}}$, the operator $\mathcal{O}_{T_L}$ is less favored, and the operator $\mathcal{O}_{S_R}$ is disfavored at $2\sigma$. 

Our analysis has been done under the assumption of one-operator dominance. However, the large observable set has revealed unique behaviors of several observables and provided many correlations between them, which allows one to distinguish between NP operators. Our analysis can serve as a map for setting up various strategies to identify the origins of NP, one of which is as follows: first, one uses the null tests $W_{IT}(q^2) = 0$ and $H_T^{(1)}(q^2) -1 = 0$ to probe the operators $\mathcal{O}_{V_R}$ and $\mathcal{O}_{T_L}$, respectively. Second, one measures the forward-backward asymmetry in $\bar{B}^0 \to D\tau^-\bar\nu_{\tau}$. If $\mathcal{A}_{FB}^D(q^2)$ has a zero-crossing point, then it is a clear sign of $\mathcal{O}_{S_L}$. The coupling $V_L$ is more difficult to test because it is just a multiplier of the SM operator. However, if the tests above disconfirm $\mathcal{O}_{V_R}$, $\mathcal{O}_{T_L}$, and $\mathcal{O}_{S_L}$ at the same time, then the modification of $V_L$ to $R(D)$ and $R(D^\ast)$ is a must. In the future when more precise data will be collected, one can adopt the strategies described here as a useful tool to discover NP in these decays if the deviation from the SM still remains. 
%\newpage
\section*{Acknowledgments}
This work was supported in part by the Heisenberg-Landau Grant and Mainz Institute for Theoretical Physics (MITP).

% ****************************************************************************
% BIBLIOGRAPHY AREA
% ****************************************************************************
%\newpage
\begin{footnotesize}
% IF YOU DO NOT USE BIBTEX, USE THE FOLLOWING SAMPLE SCHEME FOR THE REFERENCES
% ----------------------------------------------------------------------------

% ----------------------------------------------------------------------------

% IF YOU USE BIBTEX,
% - DELETE THE TEXT BETWEEN THE TWO ABOVE DASHED LINES
% - UNCOMMENT THE NEXT TWO LINES AND REPLACE 'Name_Of_Your_BibFile'

%\bibliographystyle{unsrt}
%\bibliography{Name_Of_Your_BibFile}
% example of Name_Of_Your_BibFile.bib
% @Article{Turcato:2006ch,
%      author    = "Turcato, M.",
%  collaboration = "ZEUS and H1",
%      title     = "Lepton flavour violation and charmonium physics at HERA",
%      journal   = "Nucl. Phys. Proc. Suppl.",
%      volume    = "162",
%      year      = "2006", 
%      pages     = "283-287",
%      SLACcitation  = "%%CITATION = NUPHZ,162,283;%%"
% }
% 
% @Unpublished{Gogitidze:2007du,
%      author    = "Gogitidze, N.",
%  collaboration = "H1", 
%      title     = "Prompt photons and particle momentum distributions at
%                   HERA", 
%      year      = "2007",
%      note    = "hep-ex/0701033",
%      SLACcitation  = "%%CITATION = HEP-EX 0701033;%%"
% }

\end{footnotesize}

% ****************************************************************************
% END OF BIBLIOGRAPHY AREA
% ****************************************************************************

\end{document}